\documentclass[
aps,
prb,
twocolumn,
superscriptaddress,
floatfix,
10pt
]{revtex4-2}
\usepackage[table,x11names,dvipsnames]{xcolor}
\usepackage{amsmath,amsfonts,amssymb}
\usepackage{graphicx}
\usepackage{mathtools}
\usepackage{booktabs}
\usepackage{soul}
\usepackage{color, colortbl}
\usepackage{adjustbox}
\usepackage{tcolorbox}
\usepackage{etoolbox}
\usepackage[version=4]{mhchem}
\usepackage{float}
\usepackage{listings}
\usepackage{newfloat}
\usepackage{hyperref}

\DeclareFloatingEnvironment[
    fileext=lol,
    listname={List of Listings},
    name=Listing
]{listing}

\definecolor{codegreen}{RGB}{0,128,0}
\definecolor{codeblue}{RGB}{0,0,180}
\definecolor{codegray}{RGB}{245,245,245}

\lstdefinelanguage{myyaml}{
    keywords={true,false,null,yes,no,on,off},
    sensitive=false,
    comment=[l]{\#},
    basicstyle=\ttfamily\small,
    keywordstyle=\color{blue},
    stringstyle=\color{red},
    showstringspaces=false,
    breaklines=true,
    columns=fullflexible,
    keepspaces=true
    morestring=[b]",
    morestring=[b]',
    commentstyle=\color{gray}\itshape
}

\lstdefinelanguage{yaml1}{
    language=myyaml,
    emph={
        predefined,
        prepath,
        air,
        Ag,
        Au,
        prefix,
        folder,
        subfolders,
        grating,
        sys1,
        sys2,
        sys3,
        prepend,
        eps,
        epsA,
        epsB,
        d,
        fg
    },
    emphstyle=\color{codegreen}\bfseries
}

\lstdefinelanguage{yaml2}{
    language=myyaml,
    emph={
        predefined,
        prepath,
        SiO2,
        prefix,
        folder,
        subfolders,
        sys1,
        sys3,
        prepend,
        eps,
        epsA,
        epsB,
        d,
        fg,
    },
    emphstyle=\color{codegreen}\bfseries
}

\lstdefinelanguage{mypython}{
    language=Python,
    basicstyle=\ttfamily\small,
    stringstyle=\color{red},
    emph={from,import},
    emphstyle=\color{codegreen}\bfseries,
    emph={[2]emerald},
    emphstyle={[2]\color{codeblue}},
}

\hypersetup{
	colorlinks,
	citecolor=blue,
	filecolor=black,
	linkcolor=blue,
	urlcolor=blue
}
\let\vec\mathbf
\renewcommand{\epsilon}{\varepsilon}
\newcommand{\ep}[2]{e^{i\mathbf{#1}\cdot\mathbf{#2}}}

\newcommand{\epsconv}{\hat{\epsilon}}
\newcommand{\emerald}{\textsc{\small AFLOW\nobreakdash-EMERALD}}

\newcommand{\epsconvi}{\epsconv^{-1}}
\newcommand{\zn}{\tilde{z}}
\newcommand{\zt}{\zn}

\newcommand{\hx}{H_x}
\newcommand{\hy}{H_y}
\newcommand{\hz}{H_z}
\newcommand{\ex}{E_x}
\newcommand{\ey}{E_y}
\newcommand{\ez}{E_z}
\newcommand{\kx}{\vec{K}_x}
\newcommand{\ky}{\vec{K}_y}
\newcommand{\kz}{\vec{K}_z}
\newcommand{\ki}{\vec{K}_i}

\newcommand{\bzn}{\tilde{\beta_z}}

\newcommand{\gzn}{\widetilde{\vec{G}}_z}
\newcommand{\kxn}{\widetilde{\vec{K}}_x}

\newcommand{\kin}{\widetilde{\vec{K}}_i}

\begin{document}

\title{{\sc AFLOW-EMERALD:} ElectroMagnetic modes EngineeRing in Advanced LayereD materials}

\author{Stefano Campanaro}
\email{stefano.campanaro@unimore.it}
\affiliation{Dipartimento di Fisica, Informatica e Matematica, Università di Modena e Reggio Emilia, Via Campi 213A, I-41125 Modena, Italy}
\affiliation{Department of Mechanical Engineering and Materials Science, Duke University, Durham, NC 27708, USA}

\author{Luca Bursi}
\email{luca.bursi@unimore.it}
\affiliation{Dipartimento di Fisica, Informatica e Matematica, Università di Modena e Reggio Emilia, Via Campi 213A, I-41125 Modena, Italy}
\affiliation{CNR-NANO Istituto Nanoscienze, Centro S3, Via Campi 213A, 41125 Modena, Italy}

\author{Nicholas H. Anderson}
\email{nicholas.anderson@duke.edu}
\affiliation{Department of Mechanical Engineering and Materials Science, Duke University, Durham, NC 27708, USA}
\affiliation{Center for Extreme Materials, Duke University, Durham, NC 27708, USA}

\author{Stefano Curtarolo}
\email{stefano@duke.edu}
\affiliation{Department of Mechanical Engineering and Materials Science, Duke University, Durham, NC 27708, USA}
\affiliation{Center for Extreme Materials, Duke University, Durham, NC 27708, USA}

\author{Arrigo Calzolari}
\email{arrigo.calzolari@nano.cnr.it}
\affiliation{CNR-NANO Istituto Nanoscienze, Centro S3, Via Campi 213A, 41125 Modena, Italy}
\affiliation{Department of Mechanical Engineering and Materials Science, Duke University, Durham, NC 27708, USA}
\affiliation{Center for Extreme Materials, Duke University, Durham, NC 27708, USA}

\begin{abstract}
Layered and periodically patterned heterostructures underpin advanced optical, photonic, and plasmonic (meta)materials, whose rational design demands electromagnetic solvers that are both numerically robust and tightly linked to the underlying material properties. Here, we present AFLOW-EMERALD (ElectroMagnetic modes EngineeRing in Advanced LayereD materials), an open-source, modular, Python-based computational framework for simulating electromagnetic wave propagation in finite and periodic layered (meta)materials. Built around a unified object-oriented architecture, AFLOW-EMERALD combines a numerically stable scattering-matrix method with plane-wave expansion and extends to rigorous coupled-wave analysis for laterally patterned structures such as gratings. The software computes optical spectra, spatial field distributions, and photonic band structures, including complex-k dispersion in lossy, dispersive media. A streamlined YAML workflow allows users to seamlessly import dielectric function datasets from experimental, literature, or first-principles sources. Owing to its modular design, AFLOW-EMERALD is readily extensible and suitable for integration into computational materials-design pipelines, providing a practical platform for the coupled material–geometry engineering of dielectric photonic crystals, plasmonic multilayers, hyperbolic metamaterials, and more complex architectures supporting, e.g., surface and volume plasmon-polariton modes.
\end{abstract}

\maketitle

\section{Introduction}
Layered metamaterials provide a versatile platform for engineering optical anisotropy, field confinement, dispersion, and plasmonic response across a broad spectral range~\cite{Shamim2024,Lee2024}, enabling the realization of surface and volume polariton modes~\cite{moradi2023,vpp_hmm}. A quantitative understanding of their electromagnetic (EM) behavior is therefore essential for the design and optimization of advanced optical and photonic devices.
In realistic architectures, EM propagation is governed by the interplay between material dispersion, finite stack thickness, and coupling to the external environment through interfaces or patterned layers. Photonic crystals~\cite{Joannopoulos2008}, superlattices~\cite{APELL199797}, or hyperbolic metamaterials (HMMs)~\cite{Smolyaninov2018,pod_hmm} exemplify this complexity, sustaining EM modes whose excitation critically depends on geometry, boundary conditions, and momentum matching mechanisms~\cite{excitation2014}.

Effective medium theory (EMT)~\cite{em_multilayer} offers valuable macroscopic insight into the optical response of ideal, infinite periodic systems in the deep-subwavelength limit and is widely employed. However, EMT cannot capture finite-size resonances, grating-assisted coupling, or the selective excitation of modes that emerge in realistic multilayer structures with a finite number of periods. Addressing these effects requires EM solvers capable of treating layered media beyond homogenized descriptions, while retaining a direct connection with the underlying material properties.

Here we present \emerald~(\underline{E}lectro\underline{M}agnetic modes \underline{E}nginee\underline{R}ing in \underline{A}dvanced \underline{L}ayere\underline{D} materials), an open-source, modular, Python-based computational tool for the quantitative simulation of EM wave propagation in finite and infinite layered (meta)materials and the analysis of the supported eigenmodes. The code is designed around an object-oriented architecture that promotes flexibility and scalability, while maintaining computational efficiency. Based on a comprehensive  EM framework, \emerald~combines a numerically stable scattering-matrix method (SMM)~\cite{smm1,smm2} for finite multilayer systems with plane-wave expansion (PWE) tools for photonic bandstructure (PBS) analysis in periodic media~\cite{DeBruijn2025,Li2003}.

The scattering-matrix formalism~\cite{born-wolf1999} enables the accurate computation of reflectance, transmittance, absorptance, and EM field distributions~\cite{berreman, smm2}. When applied to finite stacks, the SMM inherently captures interface reflections, grating-assisted momentum matching, and coupling to the external environment, thereby providing a realistic description of wave propagation through layered (meta)materials and a direct framework for comparison with experimental spectra. By combining the SMM with plane-wave expansions of the EM fields and dielectric functions, this approach naturally extends to laterally periodic structures, forming the basis of rigorous coupled-wave analysis (RCWA)~\cite{Moharam1981}. In parallel, the PWE formalism provides access to the Bloch eigenmodes and PBS of infinite periodic systems~\cite{Li2003,Joannopoulos2008}. This dual framework establishes a direct link between the propagating modes of finite stacks and the dispersion relation of an {\em ideal} infinite  layered material. Consequently, it allows one to identify which photonic modes can be excited under the boundary conditions dictated by the finite-thickness stack and the coupling medium, serving as a powerful tool for predicting and designing layered structures with tailored optical responses.

By integrating these complementary formalisms within a single framework, \emerald~enables quantitative simulations of optical and plasmonic properties in complex layered architectures, including dispersive and anisotropic materials. The code has been successfully used for a broad class of systems, ranging from dielectric photonic crystals to plasmonic and hyperbolic multilayers, and to the analysis of plasmon-polariton modes in complex metamaterial platforms~\cite{HMMs_arXiv_2026}.

\begin{table}[!b!]
\begin{tabular}{ll}
\hline
\textbf{Acronym} & \textbf{Definition} \\ \hline
CSV & comma-separated value \\
EM & electromagnetic \\
EMT & effective medium theory \\
HMM  & hyperbolic metamaterial\\
PBS & photonic bandstructure \\
PWE & plane-wave expansion \\
QE & quantum ESPRESSO \\
RCWA & rigorous coupled-wave analysis \\
RTA & reflectance, transmittance, \\
 & absorptance\\
SMM & scattering-matrix method \\
SPP & surface plasmon polariton \\
TE & transverse electric \\
TM & transverse magnetic \\
VPP & volume plasmon polariton \\
\hline
\end{tabular}
\caption{\small List of acronyms used in text.}
\label{tab:acr}
\end{table}

\emerald~is released as open-source software under the GNU General Public License v3.0 or later (GPL-3.0-or-later), ensuring that the code can be used, modified, and redistributed while preserving the same freedoms in distributed derivative versions. \emerald~is part of the \textsc{\small AFLOW} ecosystem for computational  design and characterization of complex materials and superstructures~\cite{nmatHT,aflow4,curtarolo:art115}.
The source code is publicly available at \url{https://github.com/aflow-org/emerald}. The repository includes the source code, documentation, example input files, and scripts required to reproduce the simulations reported in this work. We foresee a closer integration of \emerald~into the \textsc{\small AFLOW} suite, to perform thermodynamic and kinetic analysis of devices based on ultra-high-temperature plasmonic ceramics~\cite{curtarolo:art187,curtarolo:art223,deed}.

The article contains several acronyms. To facilitate reading, we add Table \ref{tab:acr}.

\section{Theoretical framework}
\emerald~provides the EM response of layered, periodically structured media by solving Maxwell's equations in the frequency domain. The implementation is based on the RCWA approach, which exploits the spatial periodicity of the system by expanding both the EM fields and the dielectric permittivity as Fourier series~\cite{Moharam1981,Moharam95,Li96}. Within each layer, decomposing the EM fields into a finite set of spatial harmonics (Floquet--Bloch modes), RCWA reduces Maxwell's equations to a matrix eigenvalue problem. The resulting eigenvectors represent the propagating and evanescent coupled waves that characterize the scattered EM states. By enforcing the continuity of the tangential field at each interface and propagating these modes through the multilayer stack, the framework yields the complete EM response as a function of wavelength, angle of incidence, and polarization. This formulation is particularly suited to photonic crystals and to metamaterial systems incorporating patterned couplers such as gratings, where periodic modulations provide the reciprocal lattice vectors required for momentum matching to free space.

\begin{figure}[t]
\centering
\includegraphics[width=0.85\linewidth]{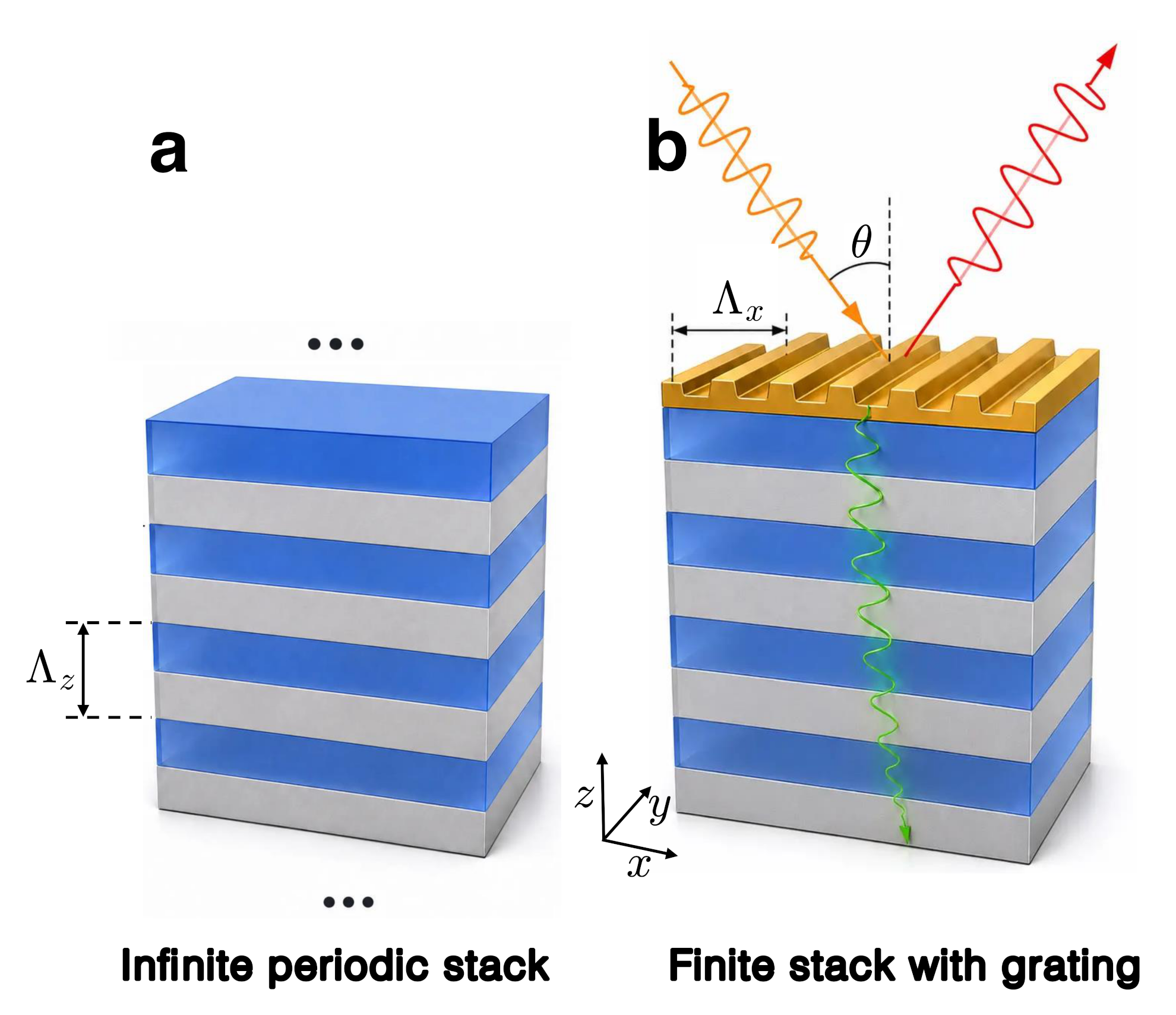}
\caption{\small Schematics of stacked planar multilayer structures. {\bf(a)} Fully periodic, three-dimensional supercrystal. {\bf(b)} Finite-thickness multilayer stack comprising a finite number of layers, optionally incorporating a grating coupler.}
\label{fig:fig1}
\end{figure}

\subsection{Maxwell's equations and conventions}
\label{sec:MaxwellConv}
We consider linear, non-magnetic media ($\mu=\mu_0$)
 with a spatially varying, dispersive complex dielectric function $\epsilon(\mathbf{r})$.
 The model geometry consists of planar layers stacked along $z$, each periodic in the $x$--$y$ plane and homogeneous along the perpendicular direction. Two configurations are treated (Figure~\ref{fig:fig1}): fully periodic multilayers, constituting three-dimensional supercrystals (panel a), and finite-thickness multilayer stacks composed of a finite number of layers (panel b). The in-plane periodicity allows the dielectric function to be expanded as a Fourier series over reciprocal lattice vectors $\mathbf{G}$,
\begin{equation}
\epsilon(\mathbf{r}) = \sum_{\mathbf{G}} \epsilon(\mathbf{G},z)\, e^{i \mathbf{G}\cdot\mathbf{r}},
\label{eq:1}
\end{equation}
where $\mathbf{G} = 2\pi \left( \displaystyle \frac{n_x}{L_x}, \frac{n_y}{L_y}\right)$ with $L_x$, $L_y$ being the periodic pitch along  $x$ and $y$, and $(n_x,n_y)\in\mathbb{Z}$.

According to Bloch's theorem, the EM fields can be written as
\begin{equation}
\mathbf{E}(\mathbf{r}) = \sum_{\mathbf{G}} \mathbf{E}(\mathbf{G},z)\ep{K_G}{r}
, \qquad
\mathbf{H}(\mathbf{r}) = \sum_{\mathbf{G}} \mathbf{H}(\mathbf{G},z)\ep{K_G}{r}
, \label{eq:bloch_fields}
\end{equation}
where $\vec{K}_\vec{G}=\beta-\vec{G}$ labels the plane-wave components and $\boldsymbol{\beta}$ is the Bloch wavevector in the first Brillouin zone.

Assuming harmonic time-dependence of the form
$\vec{E}(\vec{r}, t) = \Re\left[\vec{E}(\vec{r}) e^{-i \omega t}\right]$ and $\vec{H}(\vec{r}, t) = \Re\left[\vec{H}(\vec{r}) e^{-i \omega t}\right]$
for the EM fields, Maxwell's curl equations read
\begin{equation}
\begin{aligned}
\nabla \times \mathbf{E}(\mathbf{r}) = i \omega \mu_0 \mathbf{H}(\mathbf{r}), \qquad
\nabla \times \mathbf{H}(\mathbf{r}) = - i \omega \epsilon_0\epsilon(\mathbf{r}) \mathbf{E}(\mathbf{r}).
\end{aligned}
\label{eq:maxwell_freq}
\end{equation}

In Fourier space, the product $\epsilon(\mathbf{r})\mathbf{E}(\mathbf{r})$ gives rise to a convolution over reciprocal lattice vectors,
\begin{equation}
	\begin{aligned}
		\epsilon(\vec{r}) \vec{E}(\vec{r}) &=  \sum_{j} \left( \sum_{i} \epsilon(\vec{G}_j- \vec{G}_i,z) \vec{E}(\vec{G}_i,z) \right) e^{i \vec{G}_j \cdot \vec{r}} ,\\
	\end{aligned}
\end{equation}
that naturally leads to the definition of a permittivity convolution matrix $\hat{\epsilon}=\{\epsilon\}_{ij}$ with
$\epsilon_{ij}  =\epsilon (\mathbf{G}_i-\mathbf{G}_j, z, \omega)$ and
where the dependence on frequency $\omega$ is also made explicit.
This representation casts Maxwell's equations into a linear algebra problem for the Fourier coefficients of the fields, solvable numerically upon truncation of the plane-wave basis set. For the periodic directions, the $\nabla$ operator acting on Bloch-periodic fields reduces to algebraic multiplications in reciprocal space, while the derivative along the propagation direction must be treated explicitly $\nabla \rightarrow  [i(K_x,K_y), \partial_z]$. For fully periodic systems, the
$z$-dependence of Eqs.~(\ref{eq:1}--\ref{eq:maxwell_freq}) is likewise replaced by the corresponding Fourier components, and the $\nabla$ operator reduces to a fully algebraic form:  $\nabla \rightarrow  i(K_x, K_y, K_z)$.
This formulation constitutes the basis for both the PWE approaches to PBS analysis and the scattering-matrix calculations implemented in \emerald.

\subsection{Photonic bandstructure}
\label{sec:pbs}
We consider a medium that is periodic in the three spatial directions (3D crystal), or in the $x$--$y$ plane (2D crystal) and homogeneous along the stacking direction $z$ in the case of layered systems.
The PWE formalism gives direct access to the Bloch eigenmodes of an \emph{infinite} periodic medium and to its PBS, which describes the EM states sustained by the material. The PBS further enables a straightforward interpretation of resonant features (such as surface and volume plasmon-polariton branches) as Bloch modes selected by momentum matching, finite-thickness quantization, and grating-assisted coupling~\cite{HMMs_arXiv_2026}.

In dispersive and absorbing media (such as plasmonic and metamaterial systems), the dielectric function is generally complex and frequency dependent, $\epsilon=\epsilon_r(\omega)+i\epsilon_i(\omega)$, where $\epsilon_r$ and $\epsilon_i$ are the real and imaginary parts, respectively. The frequency dependence of $\epsilon$ turns the algebraic solution of Eqs.~(\ref{eq:maxwell_freq}) into a nonlinear eigenvalue problem that cannot be treated within a conventional PWE approach~\cite{FigotinVitebskiy2006}.
To overcome this limitation, \emerald~adopts an inverse-dispersion (or complex-$\mathbf{k}$) formulation~\cite{FigotinVitebskiy2006, Rybin2017} that recasts the problem at fixed frequency and treats a component of the Bloch wavevector as the eigenvalue. The PBS is expressed as a map of $k_z(\omega,k_x)$  -- or equivalently $k_z(E,k_x)$ -- and EM attenuation is quantified directly through $\mathrm{Im}(k_z)$.

We start considering a structure that is invariant along $y$ (1D periodic grating along $x$). For incidence, radiation in the $x$--$z$ plane ($\beta_y=0$) one has $\mathbf{K}_y=0$, and the EM problem reduces to two uncoupled polarization eigenproblems, corresponding to \emph{transverse electric} (TE, $E_y\neq 0$) and \emph{transverse magnetic} (TM, $H_y\neq 0$) polarizations.
In the case of TE modes,  in the absence of reciprocal-lattice components along the $z$ direction, the algebraic Maxwell's equation
can be rewritten as a linear eigenvalue problem for $\beta_z$ at fixed $\omega$, which remains well defined also in the presence of material dispersion:
\begin{equation}
\begin{cases}
   \gzn\hx+i\left( \kxn^2-\hat{\epsilon}\right) \ey=\bzn\hx \\
i\vec{I}\hx+\gzn\ey=\bzn\ey.
\end{cases}
\label{eq:5}
\end{equation}
In Eq.~(\ref{eq:5}) we assumed the notation $\vec{K}_i=\beta_i\vec{I}-\vec{G}_i$, $\kin ={\ki}/{k_0}$, $\widetilde{\vec{G}}_i ={\vec{G}_i}/{k_0}$, and $\tilde{\beta_i} ={\beta_i}/{k_0} $, where $k _0\equiv \omega/c$ is the wavenumber of the radiation in vacuum.

In the general case of a full periodic system, the latter equations for TE polarization can be rewritten as a matrix equation:
\begin{equation}
\begin{pmatrix}
 	\gzn&i\left(\kxn^2-\epsconv \right)  \\
 	i\vec{I} & \gzn
 \end{pmatrix}
  \begin{pmatrix} \hx \\ \ey \end{pmatrix} =
   \bzn
 \begin{pmatrix} \hx \\ \ey \end{pmatrix};
 \label{eqme}
\end{equation}
along the same lines, the equations for TM polarization read:
\begin{equation}
 \begin{pmatrix}
	\gzn&i\kxn\epsconvi\kxn-\vec{I} \\
	i\epsconv & \gzn
\end{pmatrix}
\begin{pmatrix} \ex \\ \hy \end{pmatrix} =
\bzn
\begin{pmatrix} \ex \\ \hy \end{pmatrix}.
 \label{eqmh}
\end{equation}

In the particular case of non-dispersive materials (i.e., $\epsilon=constant$), Eqs.~(\ref{eqme}-\ref{eqmh}) reduce to a conventional (i.e., linear) PWE band-structure eigenproblem:
\begin{eqnarray}
\left( \kx^2+\kz^2\right)\ey&=&k_0^2 \epsconv\ey \qquad \text{(TE)} \phantom{a} \\
\left( \kx\epsconvi\kx+\kz\epsconvi\kz\right)\hy&=&k_0^2\hy \qquad \text{(TM)} \phantom{a}
\label{eqh}
\end{eqnarray}
where the Bloch wavevector $\boldsymbol{\beta}$ is fixed along a chosen high-symmetry path in the Brillouin zone, and the eigenvalue problem is solved for the frequency $\omega$ (or equivalently $k_0=\omega/c$).

For a given $\boldsymbol{\beta}$, the corresponding plane-wave matrices $\mathbf{K}_x$ and $\mathbf{K}_z$ are constructed from the reciprocal lattice vectors, and the eigenvalues $k_0$ are obtained by solving the appropriate polarization-resolved eigenproblem. Repeating this procedure along the selected path yields the photonic bandstructure, while the associated eigenvectors provide the Fourier amplitudes of the Bloch modes~\cite{Joannopoulos2008,JohnsonJoannopoulos2001}.

\subsection{The scattering matrix method}
We first consider the case of a single  layer homogeneous along $z$ and periodic in the $x$--$y$ plane.
This allows one to decompose the complex dielectric function in Fourier components as in Eq.~(\ref{eq:1}).
The component-wise Maxwell's equations, Eq.~(\ref{eq:maxwell_freq}), for the single layer read:
\begin{equation}
	\left\{
	\begin{aligned}
		- \frac{\partial \widehat{E}_y}{\partial z} + i \ky\ez &=k_0\hx \\
		\frac{\partial \widehat{E}_x}{\partial z} - i \kx\ez&=k_0\hy \\
		- i \ky\ex - i\kx\ey &=k_0\hz\\
		- \frac{\partial \widehat{H}_x}{\partial z} + i\ky\hz&= k_0\epsconv\ex \\
		\frac{\partial \widehat{H}_y}{\partial z} - i\kx\hz&= k_0\epsconv\ey\\
		i\ky\hx+ i \kx\hy&= k_0\epsconv\ez
	\end{aligned}
	\right.
	\label{sistema}
\end{equation}
Since the longitudinal components $\ez$ and $\hz$ can be expressed as linear combinations of in-plane field components, i.e., third and sixth lines in Eq.~(\ref{sistema}), the EM problem
can be recast in terms of the field vectors  $\vec{e}_T \equiv (\ex,\ey)^T$ and  $\vec{h}_T \equiv (\hx,\hy)^T$ transverse to the stacking direction.
In the case of the electric field vector, Eq.~(\ref{sistema}) can be reformulated as a compact second-order differential equation for $\vec{e}_T$:
\begin{equation}
\frac{d^2 \vec{e}_T}{d\zt^2} = \mathbf{P}\mathbf{Q}\,\vec{e}_T
\equiv \boldsymbol{\Omega}^2\,\vec{e}_T \:,
\label{eq:Omega_def}
\end{equation}
where $\zt \equiv k_0 z$ and  $\mathbf{P}$, $\mathbf{Q}$ are $2\times 2$ block matrices:
\begin{eqnarray}
\mathbf{P}&=&
\begin{pmatrix}
\kx\,\epsconv^{-1}\,\ky & \mathbf{I}-\kx\,\epsconv^{-1}\,\kx \\
\ky\,\epsconv^{-1}\,\ky-\mathbf{I} & -\ky\,\epsconv^{-1}\,\kx
\end{pmatrix}, \\ \nonumber
\mathbf{Q}&=&
\begin{pmatrix}
\kx\,\ky & \epsconv-\kx\,\kx \\
\ky\,\ky-\epsconv & -\ky\,\kx
\end{pmatrix}.
\label{eq:PQ_defs}
\end{eqnarray}
The matrix $\boldsymbol{\Omega}^2$ is decomposed as $\boldsymbol{\Omega}^2 = \mathbf{W}\,\boldsymbol{\Lambda}^2\,\mathbf{W}^{-1}$, where  the diagonal part  $\boldsymbol{\Lambda}$
contains complex-valued propagation constants to include evanescent modes, and $\mathbf{W}$ is the eigenvector's matrix.
A similar set of equations can be written for the transverse magnetic field vector $\vec{h}_T$.
By defining $\mathbf{V}\equiv \mathbf{Q}\mathbf{W}\boldsymbol{\Lambda}^{-1}$, the transverse fields are:
\begin{equation}
\begin{aligned}
\vec{e}_T(\zt) &= \mathbf{W}\,e^{-\boldsymbol{\Lambda}\zt}\,\vec{c}^{+}
+\mathbf{W}\,e^{+\boldsymbol{\Lambda}\zt}\,\vec{c}^{-},\\
\vec{h}_T(\zt) &= -\mathbf{V}\,e^{-\boldsymbol{\Lambda}\zt}\,\vec{c}^{+}
+\mathbf{V}\,e^{+\boldsymbol{\Lambda}\zt}\,\vec{c}^{-},
\end{aligned}
\label{eq:fields_layer}
\end{equation}
where $\vec{c}^{+}=\mathbf{W}^{-1}\vec{e}^+(0)$ and $\vec{c}^{-}=\mathbf{W}^{-1}\vec{e}^-(0)$ are the modal amplitude vectors associated with the forward- and backward-propagating EM eigenwaves within a single layer.

The extension from a single layer to a multilayer stack requires tracking the evolution of the EM fields across multiple interfaces.  \emerald~addresses the issue by
 implementing the scattering-matrix formalism, which is well established for its robustness and numerical stability in thick, lossy, or strongly mismatched stacks and in systems comprising many layers~\cite{smm1,smm2,born-wolf1999}.
\begin{figure}[t]
\centering
\includegraphics[width=0.95\linewidth]{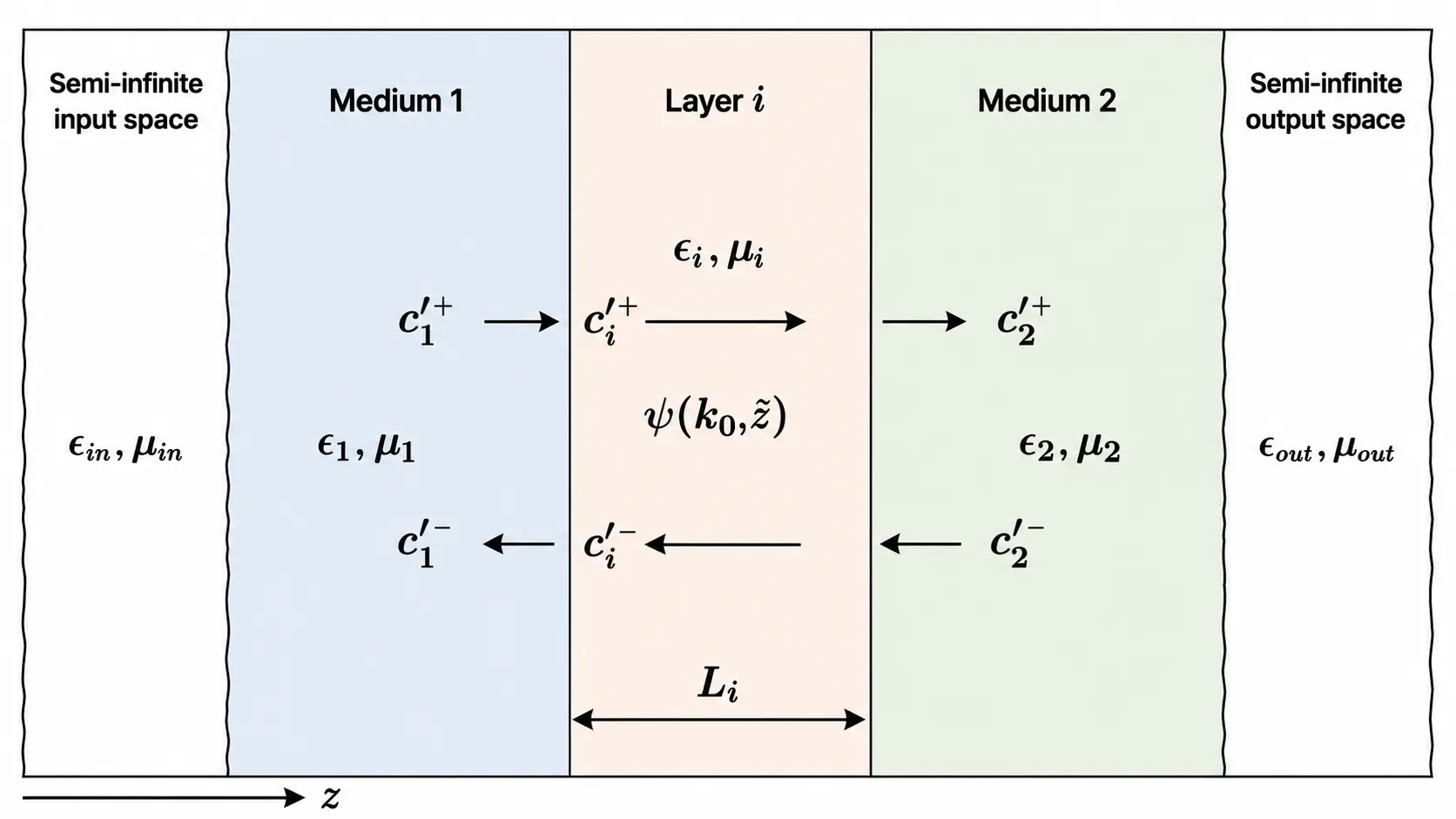}
\caption{\small Graphical representation of an EM field propagating through a planar multilayer structure aligned along the $z$-axis. $\epsilon_i$, $\mu_i$, are the electrical permittivity and magnetic permeability of  the $i$-th layer. The field within the layer is represented by $\psi_i(k_0,\tilde{z})$, while  $\mathbf{c}_i^{\pm}$  and $\mathbf{c}_i^{'\pm}$ indicate the mode coefficients inside and outside the $i$-th layer, respectively.}
\label{fig:fig2}
\end{figure}
We define a {\em two-port} scattering matrix $\mathbf{S}$ as:
\begin{equation}
\begin{pmatrix}
	\mathbf{c}_1^{'-} \\
	\mathbf{c}_2^{'+}
\end{pmatrix}
=\vec{S}
\begin{pmatrix}
	\mathbf{c}_1^{'+} \\
	\mathbf{c}_2^{'-}
\end{pmatrix} \: ,
\qquad
\mathbf{S}\equiv
\begin{pmatrix}
	\mathbf{S}_{11} & \mathbf{S}_{12} \\
	\mathbf{S}_{21} & \mathbf{S}_{22}
\end{pmatrix} \: ,
\label{eq:smatrix}
\end{equation}
where $(\mathbf{c}_1^{'+},\mathbf{c}_1^{'-})$ are incoming/outgoing modal amplitudes of the EM field, Eq.~(\ref{eq:fields_layer}), on the input side, and $(\mathbf{c}_2^{'-},\mathbf{c}_2^{'+})$ those on the output side, as shown in Figure~\ref{fig:fig2}.
Within the $i$-th layer of thickness $L_i$, we collect the transverse fields  of Eq.~\eqref{eq:fields_layer} into the state vector:
\begin{eqnarray}
\boldsymbol{\psi}_i(k_0,\zt)&=&
\begin{pmatrix}
\vec{e}_T(\zt)\\
\vec{h}_T(\zt)
\end{pmatrix}\\ \nonumber
&=&
\begin{pmatrix}
\mathbf{W}_i & \mathbf{W}_i\\
-\mathbf{V}_i & \mathbf{V}_i
\end{pmatrix}
\begin{pmatrix}
e^{-\boldsymbol{\Lambda}_i \zt} & \mathbf{0}\\
\mathbf{0} & e^{+\boldsymbol{\Lambda}_i \zt}
\end{pmatrix}
\begin{pmatrix}
\vec{c}_i^+\\
\vec{c}_i^-
\end{pmatrix}.
\label{eq:psi_layer}
\end{eqnarray}

Because of the field continuity at the interface, the scattering matrix of layer $i$  explicitly depends not only on its intrinsic properties (permittivity and thickness), but also on the dielectric functions of the adjacent regions.
To remove the dependence, we adopted the construction approach introduced in Refs.~\cite{smm1,smm2}, which inserts auxiliary zero-thickness free-space gaps between each physical layer. Each layer is thereby assigned a single, self-contained scattering matrix, independent of its position within the multilayer stack, with the added benefit of improved numerical stability and memory efficiency.
According to this construction approach, the S-matrix for  the $i$-th layer is:
\begin{equation}
\left\{
\begin{aligned}
&\mathbf{S}_{11}^{(i)}=\mathbf{S}_{22}^{(i)} \\
&= \left(\mathbf{A}_i-\mathbf{X}_i \mathbf{B}_i \mathbf{A}_i^{-1}\mathbf{X}_i \mathbf{B}_i\right)^{-1}
\left(\mathbf{X}_i \mathbf{B}_i \mathbf{A}_i^{-1}\mathbf{X}_i \mathbf{A}_i-\mathbf{B}_i\right)\\[4pt]
&\mathbf{S}_{12}^{(i)}=\mathbf{S}_{21}^{(i)} \\
&= \left(\mathbf{A}_i-\mathbf{X}_i \mathbf{B}_i \mathbf{A}_i^{-1}\mathbf{X}_i \mathbf{B}_i\right)^{-1}
\mathbf{X}_i\left(\mathbf{A}_i-\mathbf{B}_i \mathbf{A}_i^{-1}\mathbf{B}_i\right)
\label{eq:S_layer}
\end{aligned}
\right.
\end{equation}
where $\mathbf{X}_i\equiv e^{\boldsymbol{\Lambda}_i k_0 L_i}$. The auxiliary matrices $\mathbf{A}_i$ and $\mathbf{B}_i$ are defined with respect to the virtual zero-thickness gap layers (vacuum, $\epsilon_r=\mu_r=1$) inserted between adjacent physical layers~\cite{smm2}. Denoting the gap eigenvector matrices by $(\mathbf{W}_g,\mathbf{V}_g)$, we obtain:
\begin{equation}
\begin{aligned}
\mathbf{A}_i &=\mathbf{W}_i^{-1}\mathbf{W}_g+\mathbf{V}_i^{-1}\mathbf{V}_g \: , \\
\mathbf{B}_i &=\mathbf{W}_i^{-1}\mathbf{W}_g-\mathbf{V}_i^{-1}\mathbf{V}_g \: .
\end{aligned}
\label{eq:AB_gap}
\end{equation}
This formulation renders the scattering matrix of each layer symmetric and independent of its neighbors, it ensures continuity of the tangential fields at every interface, and improves numerical robustness.

To connect the finite multilayer stack to the external environment, it is interfaced with semi-infinite input/output media of permittivities $\varepsilon_{in}$ and $\varepsilon_{out}$ (Fig.~\ref{fig:fig2}), through the introduction of the input and output interface scattering matrices $\mathbf{S}^{(in)}$ and $\mathbf{S}^{(out)}$:
\begin{equation}
\begin{aligned}
\mathbf{S}^{(in)}=
\begin{pmatrix}
-\mathbf{A}_r^{-1}\mathbf{B}_r & 2\,\mathbf{A}_r^{-1}\\
\left(\mathbf{A}_r-\mathbf{D}_r\right)/2 & \mathbf{B}_r \mathbf{A}_r^{-1}
\end{pmatrix} \: , \\
\mathbf{S}^{(out)}=
\begin{pmatrix}
\mathbf{B}_t\mathbf{A}_t^{-1} & \left(\mathbf{A}_t-\mathbf{D}_t\right)/2\\
2\,\mathbf{A}_t^{-1} & -\mathbf{A}_t^{-1}\mathbf{B}_t
\end{pmatrix} \: ,
\end{aligned}
\label{eq:S_interfaces}
\end{equation}
with
\begin{equation}
\begin{aligned}
& \mathbf{D}_\alpha=\mathbf{B}_\alpha\,\mathbf{A}_\alpha^{-1}\mathbf{B}_\alpha \: , \qquad \alpha\in\{r,t\} \: , \\
& \vec{A}_\alpha=\vec{W}_g^{-1} \vec{W}_{\gamma}+\vec{V}_g^{-1} \vec{V}_{\gamma} \: , \\
& \vec{B}_\alpha=\vec{W}_g^{-1} \vec{W}_{\gamma}-\vec{V}_g^{-1} \vec{V}_{\gamma} \: , \\
\end{aligned}
\end{equation}
where $\gamma (\alpha = r) =IN$ and  $\gamma (\alpha = t) = OUT$.
The global scattering matrix is then built by cascading the individual sections through the Redheffer star product~\cite{redheffer1959} (here denoted by $\otimes$),
\begin{equation}
\mathbf{S}^{(\text{global})}=
\mathbf{S}^{(in)}\otimes
\mathbf{S}^{(\text{device})}\otimes
\mathbf{S}^{(out)}.
\label{eq:S_global}
\end{equation}

Given an incident (\verb|inc|) modal-amplitude vector $\mathbf{c}_{\verb|inc|}$,
the reflected (\verb|ref|) and transmitted (\verb|trn|) modal coefficients follow directly from Eq.~(\ref{eq:smatrix}):
\begin{equation}
\begin{aligned}
\mathbf{c}_{\texttt{ref}}&=\mathbf{S}_{11}^{(\texttt{global})}\,\mathbf{c}_{\texttt{inc}} \: , \\
\mathbf{c}_{\texttt{trn}}&=\mathbf{S}_{21}^{(\texttt{global})}\,\mathbf{c}_{\texttt{inc}} \: .
\end{aligned}
\label{eq:cref_ctrn}
\end{equation}
The corresponding transverse electric-field Fourier coefficients are
\begin{equation}
\begin{aligned}
\vec{r}_T&=\mathbf{W}_{\texttt{ref}}\,\mathbf{c}_{\texttt{ref}} \: , \\
\vec{t}_T&=\mathbf{W}_{\texttt{trn}}\,\mathbf{c}_{\texttt{trn}} \: ,
\end{aligned}
\label{eq:rT_tT}
\end{equation}
where $\mathbf{W}_{\texttt{ref}}$ and $\mathbf{W}_{\texttt{trn}}$ matrices collect the eigenvectors of the reflected and transmitted EM modes, respectively.

The total reflectance $R$ and transmittance $T$ spectra are obtained by summing over all propagating modes:
\begin{equation}
\begin{aligned}
R&=\sum_{m,n}  \Re\!\left(\frac{K_{z,mn}^{\texttt{ref}}}{K_{z}^{\texttt{inc}}}\right)
\left|\vec{r}_{mn}\right|^2\: \\
T&=\sum_{m,n} \Re\!\left(\frac{K_{z,mn}^{\texttt{trn}}}{K_{z}^{\texttt{inc}}}\right)
\left|\vec{t}_{mn}\right|^2 \: ,
\end{aligned}
\label{eq:RT_tot}
\end{equation}
where $K_{z,mn}^{\texttt{ref}}$ and $K_{z,mn}^{\texttt{trn}}$ are the longitudinal wavevector components of all possible reflected and transmitted modes, respectively, with $n,m\in\mathbb{Z}$.
The total absorptance spectrum $A$ follows from energy conservation as
$A = 1 - R - T$.
This formalism rigorously describes energy distribution among all diffraction orders, while enforcing energy conservation in lossless systems and naturally accounting for absorption in lossy media.

\section{Software Design}
\emerald~is implemented in Python 3.9 and MATLAB. The Python layer consists of a single object-oriented class and a collection of supporting functions; the key dependencies are listed in Table~\ref{tab:library}. Computationally intensive tasks are delegated to vectorized MATLAB routines, which can be compiled for GPU acceleration. The Python engine handles input parsing and orchestrates the calls to these core functions.

\begin{figure*}
	\centering
	\includegraphics[width=0.85\textwidth]{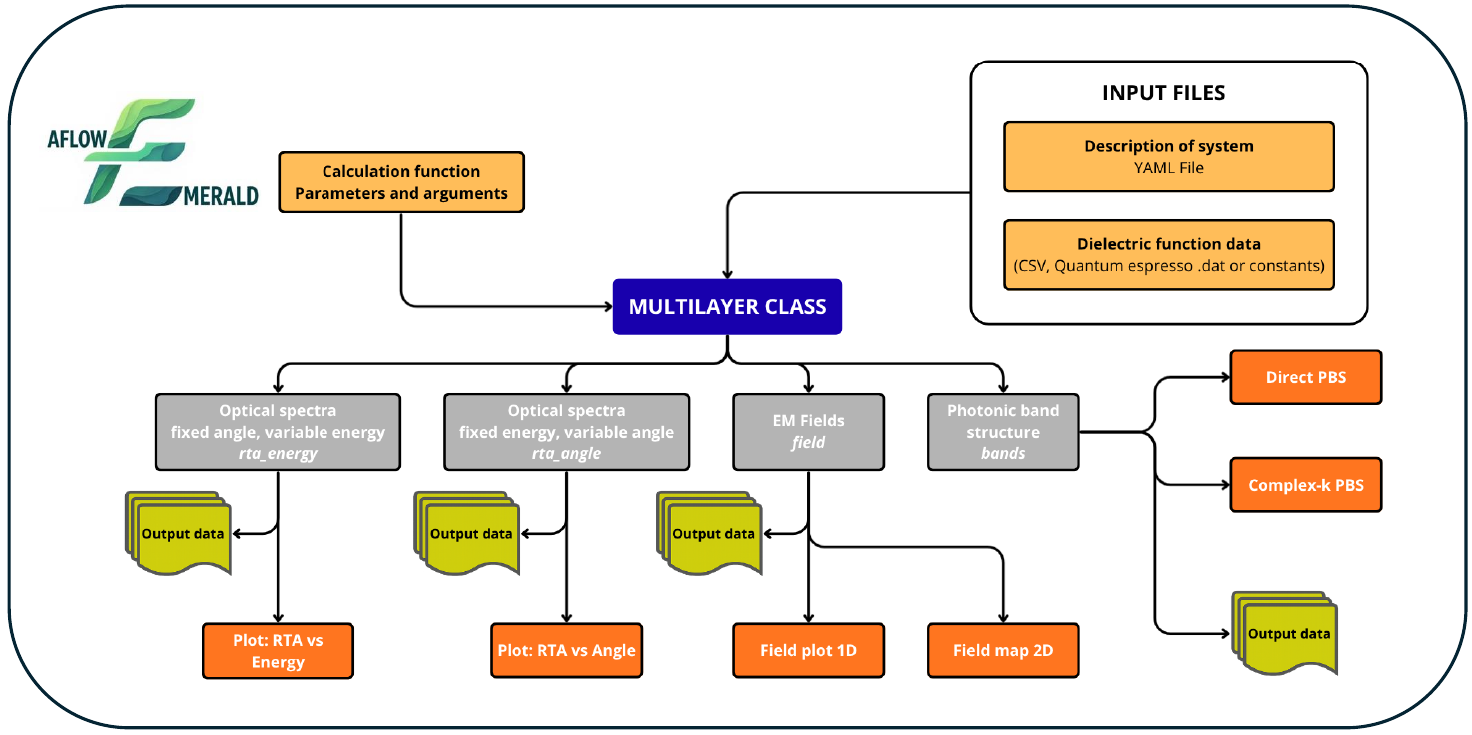}
	\caption{\small Schematic workflow of the {\small AFLOW-EMERALD} software. Input files specifying the multilayer geometry and dielectric properties are parsed by the \texttt{Multilayer} class and processed by the computational engine. The software calculates optical spectra - namely reflectance, transmittance, and absorptance  (RTA) -  electromagnetic field distributions, and photonic bandstructures using both direct and complex-$\mathbf{k}$ formulations.}
	\label{fig:fig3}
\end{figure*}

The operational workflow  of {\small AFLOW-EMERALD} software is schematized in Figure~\ref{fig:fig3}.
The software is designed to operate within a Jupyter Notebook environment, where multiple simulations can be executed and their outputs analyzed interactively. All functionalities are contained in a single file, \texttt{multilayer.py}, which defines a main class of the same name together with several auxiliary functions (see code Snippet~\ref{lis:python}). Initialization is performed in the first notebook cell; the structural properties of the system under analysis are supplied via a dedicated YAML configuration file.
A single YAML file may describe multiple systems, each identified by a unique name. Launching a simulation requires only:
\begin{listing}[h]
\begin{lstlisting}[language=mypython, caption={\small Example of \texttt{Multilayer} class initialization.}]
from emerald import multilayer
filepath = "systems.yaml"
structure = "material1"
mat = Multilayer(filepath, structure)
\end{lstlisting}
\label{lis:python}
\end{listing}

Required inputs are the dielectric functions (or static dielectric constants, for non-dispersive materials), together with the structural parameters of the multilayer, e.g., layers thicknesses and number of periods. A full description of the input parameters, code functionalities, and generated outputs is presented in the following section. \emerald~ is distributed as a standard Python package and can be installed in a conventional Python environment. Once installed, no additional configuration is required beyond the standard Python dependencies listed in Table~\ref{tab:library}. A brief README file with installation and execution instructions is available in the GitHub repository. (\url{https://github.com/aflow-org/emerald}).

\begin{table}[]
\centering
\begin{tabular}{@{}ll@{}}
\toprule
\textbf{Python library} & \textbf{Version} \\ \midrule
Numpy & 2.1 \\
Scipy & 1.14 \\
Matplotlib & 3.9 \\
PyYAML & 1.1 \\
Matlab engine & R2024A \\ \bottomrule
\end{tabular}
\caption{\small Python libraries used along with MATLAB engine version.}
\label{tab:library}
\end{table}

\section{Code Description}
\subsection{System description}
System parameters are provided through a YAML configuration file organized into two main sections (Figure~\ref{fig:fig3}). The first defines the \emph{predefined materials}, which serve as the fundamental building blocks of the multilayer. Each material can be specified in three ways (see code Snippet~\ref{es1}): {\bf i.} by assigning a complex dielectric constant; {\bf ii.} by referencing an external  file containing the full frequency-dependent dielectric function in  comma-separated values (CSV) format; or {\bf iii.} by using the output format of Quantum ESPRESSO (QE)~\cite{qe}.
If the dielectric function data reside in a directory other than that of the active notebook, a base path can be set via the \texttt{prepath} key. Backend-specific base directories are also supported: \texttt{prepath\_csv} applies exclusively to CSV-based materials, while \texttt{prepath\_qe} applies to data obtained from the \texttt{epsilon.x} code of the QE suite. Path resolution follows three conventions: an alphanumeric-initial path is interpreted as relative to the corresponding \texttt{prepath}; a \texttt{./}-prefixed path is resolved relative to the configuration file directory, ignoring \texttt{prepath}; and a \texttt{/}-prefixed path is treated as absolute.

\begin{listing}[!h]
\begin{lstlisting}[language=yaml1, escapeinside={(*@}{@*)}, caption={\small Example of YAML configuration file for predefined materials section.}]
predefined:
  prepath: data
  air: 1
  Ag: -14.4204+1.064j
  Au: gold.csv
  (*@\color{codegreen}SiO2@*): csv # Shortcut to SiO2.csv
  (*@\color{codegreen}GaAs@*): qe  # Shortcut to GaAs.dat
  (*@\color{codegreen}InAs@*):
    prefix: InAs
    folder: InAs-calc1
    subfolders: eps1, eps2
\end{lstlisting}
\label{es1}
\end{listing}

\begin{listing}[!h]
\begin{lstlisting}[language=yaml2,escapeinside={(*@}{@*)}, caption={\small Example of YAML configuration file defining multiple systems.}]
(*@\color{codegreen}grating@*):
  (*@\color{codegreen}1@*):
    d: 10
    epsA: Ag
    epsB: air
    fg: 0.5
sys1:
  prepend: grating
  (*@\color{codegreen}1@*):
    eps: Au
    d: 10
  (*@\color{codegreen}2@*):
    eps: InAs
    d: 10
(*@\color{codegreen}sys2@*):
  (*@\color{codegreen}1@*):
    eps: 5
    d: 20
  (*@\color{codegreen}2@*):
    eps: GaAs
    d: 20
sys3:
  prepend: grating, sys2
  (*@\color{codegreen}1@*):
    eps: 9
    d: 12
\end{lstlisting}
\label{es2}
\end{listing}

The second section of the YAML file defines the systems used in the simulations.
 Since the number of periods is passed as a parameter to the simulation functions, the YAML configuration describes only the structure unit cell.
Each unit cell comprises an ordered sequence of layers, each defined by its thickness and material composition.
Systems are identified by unique keys, with the constituent layers indexed by consecutive integers starting from index 1.

If a layer is homogeneous, it is described by two parameters:
\begin{itemize}
  \item \texttt{eps}: the dielectric function of the layer, which can be specified either as a numerical value (real or complex) or as a reference to a predefined material key;
  \item \texttt{d}: the layer thickness in nanometers, which can equivalently be specified using the alias \texttt{thickness}.
\end{itemize}
If a layer is laterally inhomogeneous and/or composed of two different materials, the parameter \texttt{eps} is replaced by:
\begin{itemize}
  \item \texttt{epsA}: dielectric function of material A;
  \item \texttt{epsB}: dielectric function of material B (optional; if omitted, it is assumed to be equal to unity);
  \item \texttt{fg}: horizontal filling factor of material A.
\end{itemize}

The \texttt{include} key allows a block of layers from one system to be inserted into another's unit cell.
Multiple blocks can be imported by providing a comma-separated list of keys, as shown in code Snippet~\ref{es2}.
Layers positioned outside the unit cell are defined via the \texttt{prepend} and \texttt{append} keys. Like \texttt{include},
these keys support multiple comma-separated entries.

Importantly, blocks imported through these keys contribute only their numerically indexed layers. Any further nested inclusions are ignored.
Specifically, if a block included via the unit cell contains its own \texttt{prepend} or \texttt{append} definitions, these external layers are not propagated
to the final assembled structure.

\begin{table*}[!h!]
\begin{adjustbox}{width=\textwidth,center}
\begin{tabular}{ccccccc}
\hline
\textbf{Parameter} & \textbf{Description} &\textbf{RTA}   & \textbf{EM fields} &\textbf{Photonic} & \textbf{Mandatory} &\textbf{Default} \\
                              &                                &  \textbf{spectra} &                             &\textbf{analysis}  &                               &   \\ \hline\hline
\texttt{periods}                  & number of periods  & X                        & X                         & X                        & Yes                         & - \\\hline
\texttt{energy}                   & energy range          & X                        & X                         & X                        & Yes                         & - \\\hline
\texttt{Lambda}                 & in-plane cell width  & X                        & X                          & X                        & No                          & 100~nm \\\hline
\texttt{halfnpw}                  & (half) number         & X                        & X                          & X                        & No                          & 0 \\
                              & of plane-waves        &                           &                             &                            &                               &   \\\hline
\texttt{eps\_in}                   & dielectric constant & X                        & X                          &                            & Yes                        & - \\
                               & in-layer                  &                           &                             &                            &                               &   \\\hline
\texttt{eps\_out}                 & dielectric constant & X                        & X                          &                            & Yes                        & - \\
                               & out-layer                 &                          &                             &                            &                               &   \\\hline
\texttt{extL\_dist}               & external space        &                          & X                          &                            & Yes                        & - \\
                              & in field simulation    &                          &                              &                            &                               &   \\\hline
\texttt{plot}                        & plot format               & X                      & X                           & X                        & No                          & No plot \\\hline
\texttt{file}                         & saving picture          & X                      & X                           & X                        & No                          & - \\
                              & path                         &                         &                               &                           &                                &   \\\hline
\texttt{kx}                          &  planar wavevector  &                          &                               & X                       & Yes                         & - \\
                              & array                        &                          &                               &                          &                                &   \\\hline
\texttt{lines}                      & resonance lines       &                          &                               & X                       & No                          & True \\
                              &in 2D bandplot          &                          &                               &                           &                                &   \\\hline
\texttt{resonance}             & resonances to plot   &                         &                               &  X                       & No                          & All\\ \hline\hline
\end{tabular}
\end{adjustbox}
\caption{\small Description of the parameters contained in the Python simulation dictionary.}
\label{tab:parameters}
\end{table*}

\subsection{Code Functionalities}
The \texttt{Multilayer} class provides a suite of four main functionalities - namely ~\texttt{rta\_energy}, ~\texttt{rta\_angle},  ~\texttt{fields}, and ~\texttt{bands} (Figure~\ref{fig:fig3}) - for the evaluation and analysis of optical response spectra, field distributions, and photonic bandstructures across planar multilayers. For all functionalities, the specific simulation parameters are defined via a Python dictionary. A detailed description of these keywords is summarized in Table~\ref{tab:parameters}.\\[\baselineskip]
\noindent 1)~\texttt{rta\_energy}
 computes the RTA optical spectra (i.e., reflectance, transmittance, and absorptance) through a multilayer, as a function of energy. The calculation is performed for a fixed angle of incidence, with the energy sampled over a user-defined range.
The execution of this functionality requires:  {\bf i.} the definition of the  incidence angle  of the incoming radiation relative to the optical axis of the  stack material aligned along $z$ and perpendicular to the material layers (default: $0^{\circ}$); and {\bf ii.} a \texttt{NumPy} array containing the energy values at which the calculation is performed.
The energy values provided for the calculation need not coincide with the points in the dielectric function dataset. When the input values differ, the software estimates the corresponding permittivity values via polynomial interpolation. A single energy value can be provided instead of an array; in this case, the method performs the calculation for that specific point.
\\[\baselineskip]
\noindent 2)~\texttt{rta\_angle}
computes the RTA optical spectra as a function of the angle of incidence. The energy is held constant, while the incidence angles are provided as an input \texttt{NumPy} array.
\\[\baselineskip]
\noindent 3)~\texttt{fields}
 evaluates the real-space EM field distribution across the multilayer structure. Fields are resolved for a specified system at a fixed energy and either a fixed angle of incidence or a fixed in-plane wavevector ($k_x$). This functionality provides the 1D intensity profile of the EM field across the multilayer along the $z$-axis, together with the 2D color map of the EM modes in the
$x-z$ plane. Depending on the TE or TM polarization of the incoming radiation, either the electric or the magnetic field can be evaluated. Execution of this functionality requires: {\bf i.} definition of the energy and incidence angle of the incoming radiation relative to the optical axis, aligned along the
$z$-axis (default: $0^{\circ}$); {\bf ii.} definition of the layers external to the multilayer (Figure~\ref{fig:fig2}) for evaluation of the incoming and outgoing EM fields; {\bf iii.} definition of the spatial mesh for the EM calculation.
\\[\baselineskip]
\noindent 4)~\texttt{bands}
 computes the two-dimensional PBS, yielding an \(N \times M\) matrix corresponding to \(N\) sampled energy points and \(M\) in-plane wavevectors (\(k_x\)). By returning the complex Bloch wavevector, this method allows for the simultaneous analysis of propagating and evanescent modes. For non-dispersive systems, ~\texttt{bands}  generates standard photonic bandstructure plots representing the $E(\vec{k})$ dispersion of the EM energy (or frequency) with respect to the wavevector, calculated along high-symmetry paths within the first Brillouin zone of the periodic superstructure. In this specific case, a specific functionality (\texttt{bands3D}) allows for the calculation
of three-dimensional photonic bandstructure.
For dispersive and absorbing layered materials, the bandstructure is instead rendered as a color-scaled two-dimensional map in which the intensity encodes the imaginary part of the longitudinal Bloch wavevector Im$(k_z)$.

While PBS represents the continuum of EM modes of an infinite periodic medium, finite structures impose specific selection rules. Factors such as layer thickness,  a limited number of periods, or the use of a momentum-matching coupler (such as  grating) dictate which EM modes can actually be excited. Within the PWE framework, the EM resonances in a finite-thickness multilayer occur when the Bloch wavevector satisfies the structural periodicity. For a stack of $N$ layers, the finite geometry discretizes the allowed $k_z$ values (Figure~\ref{fig:fig1}), restricting the supported Bloch waves to those fulfilling the resonance condition:
\begin{equation}
k_{z} = \frac{\ell}{2N} \frac{2\pi}{\Lambda_z}, \quad \ell = 1, \dots, N,
\label{bloch}
\end{equation}
where  $\Lambda_z$ is the thickness of the unit cell along $z$.
For fixed $N$ and $\Lambda_z$, \emerald~displays the allowed $k_z$ solutions of Eq.~(\ref{bloch}) as colored isolines superimposed to the 2D band map.
In-plane periodicity  imparts analogous constraints on the transverse wavevector. For example, a linear grating of period $\Lambda_x$
couples the incident field to a discrete set of in-plane momenta, decomposing the transmitted field into multiple diffraction orders. The integer index
$\tilde{k}_x=(k_x\Lambda_x)/2\pi=0,1,...,n$ labels the  first and higher harmonics, respectively. The intensity of each diffracted component decreases with increasing harmonic order. The allowed $\tilde{k}_x$ values are marked as vertical colored lines on the 2D bandstructure.
In the general case, resonance conditions are located at the intersections of the isolines satisfying Eq.~(\ref{bloch}) with the vertical lines representing the in-plane momenta introduced by the periodic lateral structure.

\subsection{Computational performance}
\emerald~exhibits competitive computational performance through the combination of a modular Python architecture and vectorized MATLAB kernels. For multilayer systems with $\sim$10--20 periods and a plane-wave basis of $\sim$50--100 components, optical spectra (reflectance, transmittance, and absorptance) are computed in approximately 0.03 s per energy point on a single-core CPU. Two-dimensional PBS maps, sampling $10^{2}$--$10^{3}$  points in energy and in-plane wavevector, are obtained in 10--100 minutes depending on the size of the PWE basis
and the number of CPU cores employed. The software also enables straightforward deployment on high-performance computing (HPC) clusters, allowing efficient parallel execution of computationally demanding simulations.

The scattering-matrix formalism ensures numerical stability for thick or highly dissipative systems, suppressing the exponential error growth that afflicts transfer-matrix approaches. GPU acceleration, available through compilation of the MATLAB routines, yields speedups of up to one order of magnitude, particularly for RCWA calculations with large plane-wave bases. Memory requirements remain moderate -- typically below 1--2 GB for standard simulations -- rendering the code suitable for both local workstations and high-performance computing environments, with favorable scaling as a function of basis size and system complexity.

\section{Examples}
\subsection{Optical RTA spectra}
\begin{figure}[!t!]
\centering
\includegraphics[width=0.8\linewidth]{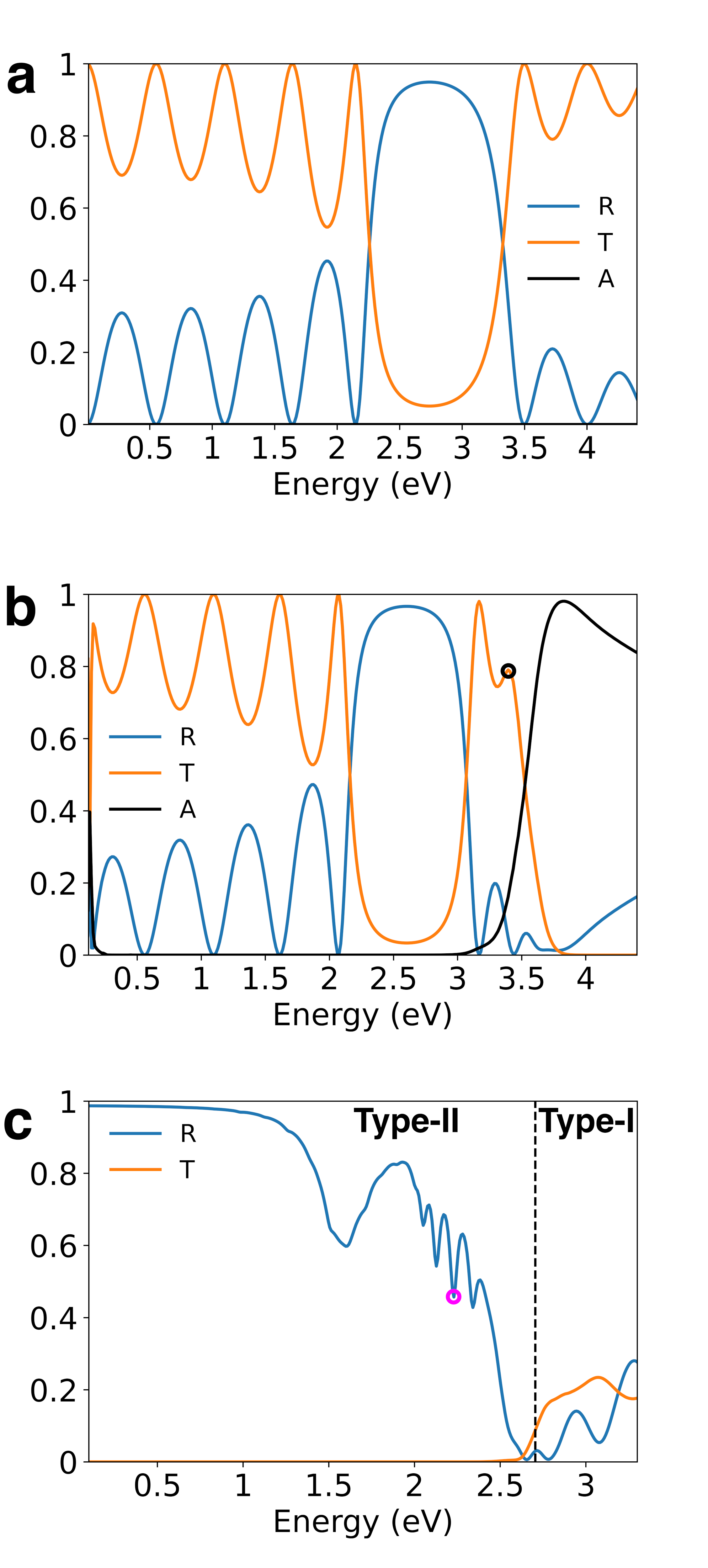}
\caption{\small Finite thickness multilayers. Optical reflectance, transmittance, and absorptance (RTA) spectra of a one-dimensional \ce{TiO2}/\ce{SiO2} photonic crystal consisting of five periods with equal layer thicknesses of 60~nm, calculated using {\bf(a)} constant real dielectric permittivities and {\bf(b)} frequency-dependent dielectric functions from reference literature~\cite{TiO2,SiO2}. {\bf(c)} Optical reflectance and transmittance spectra of Ag/\ce{TiO2} hyperbolic metamaterial composed of 6 periods with equal layer thicknesses of 25~nm. A 15~nm-thick rectangular silver grating with lateral periodicity $\Lambda_x=100$~nm is added on the stack top. Vertical dashed line identifies the type-I and type-II hyperbolic spectral ranges. Colored circles in panels b and c indicate selected EM modes analyzed in Section~\ref{sec:fields}.}
\label{fig:fig4}
\end{figure}

We applied the \texttt{rta\_energy} functionality of \emerald~to calculate the optical reflectance, transmittance, and absorptance (RTA) spectra of a prototypical finite-size photonic crystal composed of alternating \ce{TiO2} and \ce{SiO2} dielectric layers. Each layer was 60~nm thick, and the structure consisted of 5 periods embedded in air ($\epsilon_{in}=\epsilon_{out}=1$). For the dielectric functions of the two oxide materials, we considered two cases: {\bf i.}
real dielectric  constants (i.e., $\epsilon_2=0$)  fixed to their average values in the visible and near-infrared spectral range, where the variations of the dielectric functions are relatively small; {\bf ii.}  complex frequency-dependent dielectric functions (i.e., dispersive materials) extracted from literature datasets~\cite{TiO2,SiO2} and converted into CSV format.
The RTA spectra of the TiO$_2$/SiO$_2$ stack were computed over an energy range 0--4.5~eV and are displayed in Figure~\ref{fig:fig4}. TiO$_2$ and SiO$_2$ are wide-bandgap dielectrics with bandgap energies of $E_g$(TiO$_2$)=3.2~eV and $E_g$(SiO$_2$)=8.9~eV, respectively. When paired, they form a type-I band alignment, where the bandgap of TiO$_2$ is entirely nested within that of SiO$_2$. Consequently, for photon energies below $E_g$(TiO$_2$), interband transitions are forbidden (i.e., absorption is absent, $\epsilon_2=0$), and the incident EM radiation is split solely between the reflected and transmitted components. In this energy range, the reflectance and transmittance spectra obtained with constant (Fig.~\ref{fig:fig4}a) and frequency-dependent (Fig.~\ref{fig:fig4}b) dielectric functions are almost identical, with minor differences due to the numerical discrepancies in the corresponding permittivity values. For photon energies above $E_g$(TiO$_2$), interband transitions are active (i.e., $\epsilon_2 \neq 0$), leading to the emergence of a finite absorptance signal (black line in Fig.~\ref{fig:fig4}b), visible only when considering frequency-dependent permittivities, which progressively becomes the dominant spectral component of the optical response.

The same approach can be applied to multilayers combining dielectric and metallic layers.
In general, the presence of metallic components prevents EM waves from penetrating and propagating through the multilayer, causing the majority of the incident radiation to be reflected at the surface. However, under specific conditions, such as those leading to hyperbolic dispersion~\cite{ferrari},
extraordinary EM modes -- e.g., volume plasmon polaritons (VPPs) -- can propagate along sharp cones across the stack. This gives rise to non-zero transmittance and absorptance contributions (see, e.g., Ref.~\cite{HMMs_arXiv_2026}).

By employing the \emerald~software, we calculated the RTA spectra for a HMM consisting of 6 periods of Ag/TiO$_2$ alternating layers, each with an equal thickness of 25~nm.
A silver grating layer with periodicity $\Lambda_x= 100$~nm is included on the top of the stack (Figure~\ref{fig:fig1}),  to compensate the momentum mismatch between the incident radiation and the VPP~\cite{ferrari}.
The dielectric functions used as input for Ag and \ce{TiO2} have been extracted from experimental data sets  reported in Refs.~\cite{Ag-CIESIELSKI2017} and~\cite{TiO2}, respectively.
Silver serves as plasmonic material, characterized by a plasma energy of $E_p=3.8$~eV.
The multilayer structure exhibits a hyperbolic character for $E < E_p$.
Specifically, within the  energy range $[0-2.6]$~eV, the system displays type-II behavior, whereas it transitions to a type-I character in the $[2.6-3.8]$~eV range.
For $E > E_p$, the system acts as a conventional dielectric.
The results are reported in Figure~\ref{fig:fig4}c.
In the type-II hyperbolic regime, the reflectance is close to unity for $E\lesssim1.0$~eV, followed by a series of dips of decreasing intensities.
The lowest energy  minimum at $\sim 1.5$~eV corresponds to a surface-bound EM mode, whereas the subsequent minima are associated with the excitation of VPP modes propagating across the stack.
Conversely, the transmittance remains negligible over the same spectral range. This behavior indicates that, while VPP modes are successfully excited within the bulk of the metamaterial, the boundary conditions at the lower external interface prevent the EM waves from radiating into the outer medium. To enhance transmittance, appropriate grating coupling conditions could be engineered on the exit surface. A slight increase in transmittance is observed within the type-I energy range, attributed to the enhanced dielectric-like character of the stack. A rigorous classification of the EM nature of these minima is provided by the field profile analysis discussed below (Section~\ref{sec:fields}).

\subsection{Angle-resolved spectroscopy}\label{sez:angular}

\begin{figure}[!h!]
\centering
\includegraphics[width=0.8\linewidth]{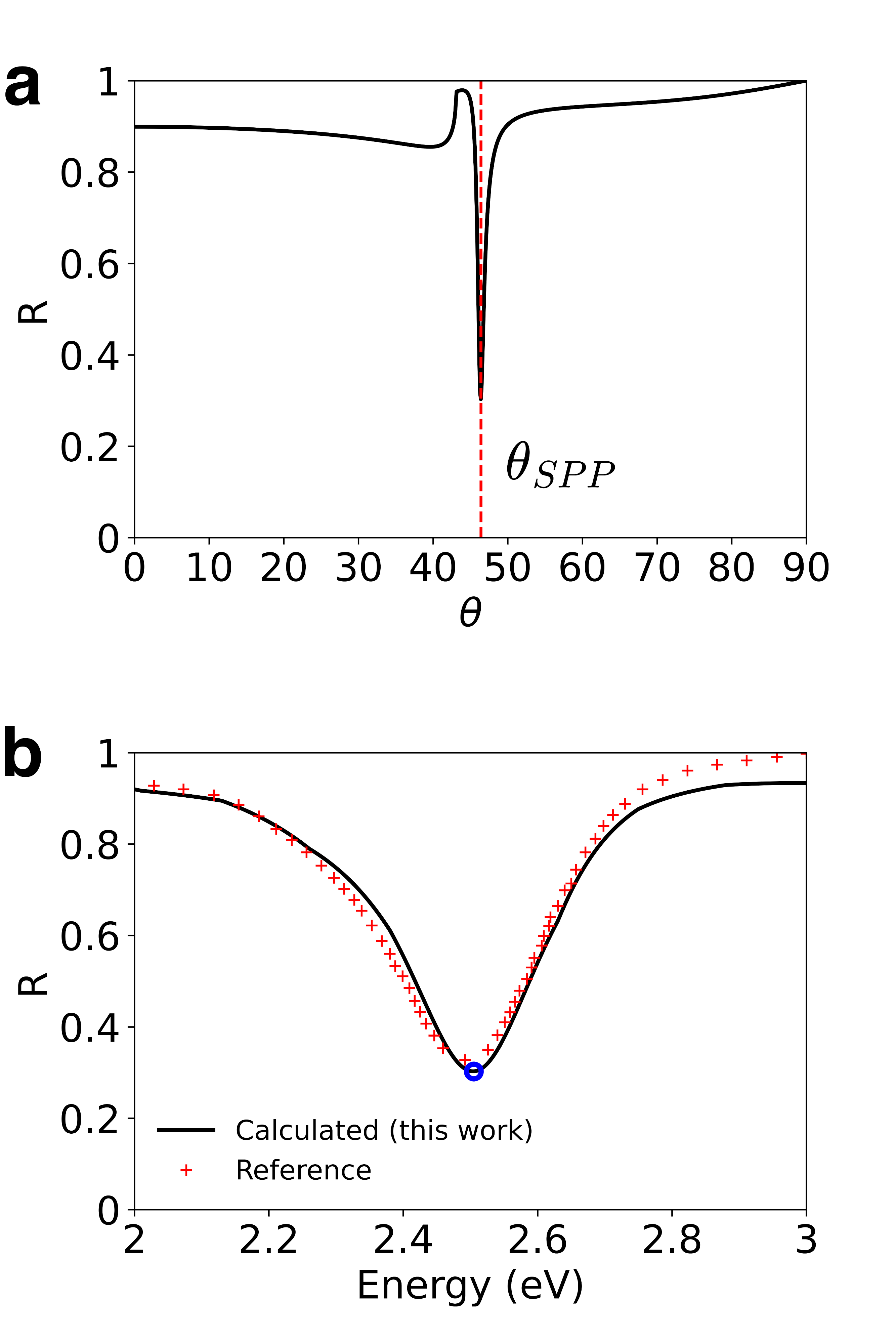}
\caption{\small {\bf(a)} Reflectance spectrum (R) of the air/Ag interface as a function
of the photon incidence angle, calculated at a fixed photon energy of $E = 2.5$~eV.
The pronounced reflectance dip identifies the resonant excitation of a surface plasmon polariton.
{\bf(b)} Energy dependent reflectance spectrum calculated with {\small AFLOW-EMERALD} (solid black line) at an incidence angle  $\theta_{SPP}=46.4^{\circ}$, corresponding to the reflectance minimum in panel a.
Experimental data from Ref.~\cite{takagi2017}  are shown as red crosses for comparison. Colored circle in panel b indicates the selected EM mode analyzed in Section~\ref{sec:fields}.}
\label{fig:fig5}
\end{figure}

As an example of the \texttt{rta\_angle} functionality, we considered the case of a surface plasmon polariton (SPP) propagating along the interface between a dielectric and a plasmonic material. Because the SPP is bound to the surface and decays in the direction perpendicular to the interface, the transmittance across the multilayer is close to zero. From the spectroscopic standpoint, the excitation of an SPP corresponds to a minimum in the reflectance, due to light absorption and plasmon excitation.
The $\mathbf{k}$-matching condition at the interface, on the other hand, selects specific angles of the incoming radiation to excite the SPP mode. The reflectance as a function of angle is therefore a suitable tool to identify the presence of SPPs. We address this with the \texttt{rta\_angle} functionality of \emerald.

We considered the testbed case of an air/Ag interface.
We calculated the reflectance spectrum of a 41~nm thick Ag film exposed to air as a function of the angle of incidence, spanning from $0^{\circ}$ (normal incidence) to $90^{\circ}$ (grazing incidence). The photon energy of the incoming radiation was fixed at $E= 2.5$~eV, which lies below the upper bound $E_{SPP}^{max}=E_p^{Ag}/ \sqrt{(1+\epsilon_1^{air})}=2.7$~eV, required to excite a SPP at the air/Ag interface.
The momentum matching is realized through a  UV fused quartz prism with refractive index $n=1.47$ in the Kretschmann configuration. The dielectric function of silver used in these simulations is taken from the experimental optical dielectric functions reported in Ref.~\cite{Ag-Johnson}.

The results, shown in Figure~\ref{fig:fig5}a, indicate a sharp minimum in the reflectance spectrum
at $\theta_{SPP}=46.4^{\circ}$  that corresponds to the excitation of the SPP.
Notably, this angle  exactly matches the experimental value for the same metal/air interface~\cite{takagi2017}.
To further compare our results with literature data, we calculated the reflectance for the air/Ag interface as a function of energy, for incoming radiation at the angle $\theta_{SPP}$ corresponding to the reflectance minimum in Figure~\ref{fig:fig5}a.  The results, shown in Figure~\ref{fig:fig5}b alongside reference experimental data~\cite{takagi2017}, clearly demonstrate the accuracy and the predictive power of the \emerald~code.
The characteristic surface-bound behavior of SPP is confirmed by the magnetic-field profile (see below).

\begin{figure*}[!h!]
\centering
\includegraphics[width=0.9\linewidth]{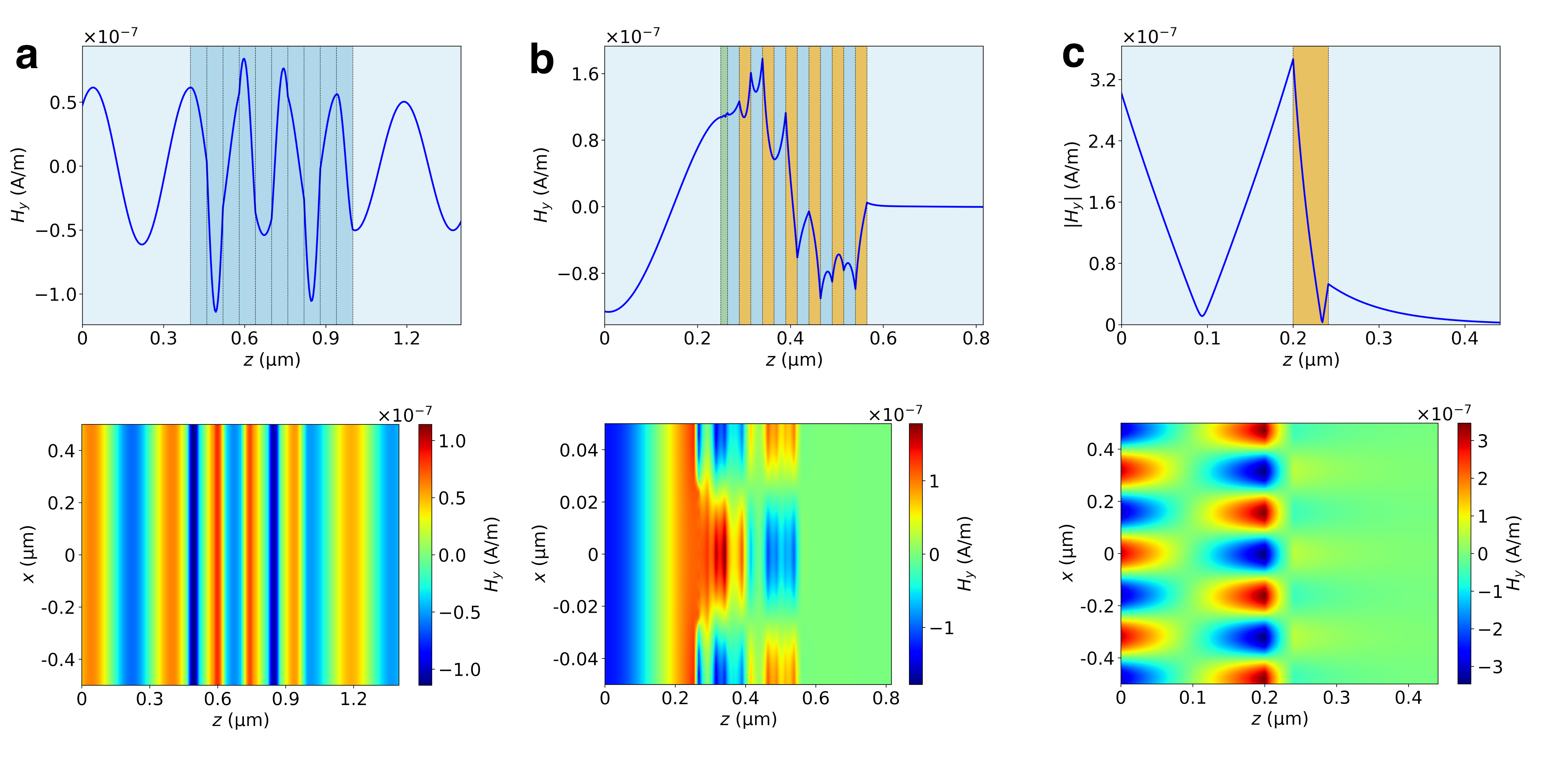}
\caption{\small Magnetic field intensity (top panels) and corresponding 2D spatial profiles (bottom panels) for selected EM modes: {\bf(a)} photonic resonance at 3.43~eV in a \ce{TiO2}/\ce{SiO2} photonic crystal; {\bf(b)} VPP resonance  at 2.23~eV in a Ag/\ce{TiO2} hyperbolic metamaterial.The green layer at the top of the system in the field plot represents the grating coupler; and {\bf(c)} SPP resonance at 2.5~eV in a air/Ag interface. The respective modes are marked by open circles in the RTA spectra of Figures~\ref{fig:fig4} [for a and b] and~\ref{fig:fig5} [for c].}
\label{fig:fig6}
\end{figure*}
\begin{figure*}[!h!]
	\centering
	\includegraphics[width=0.8\textwidth]{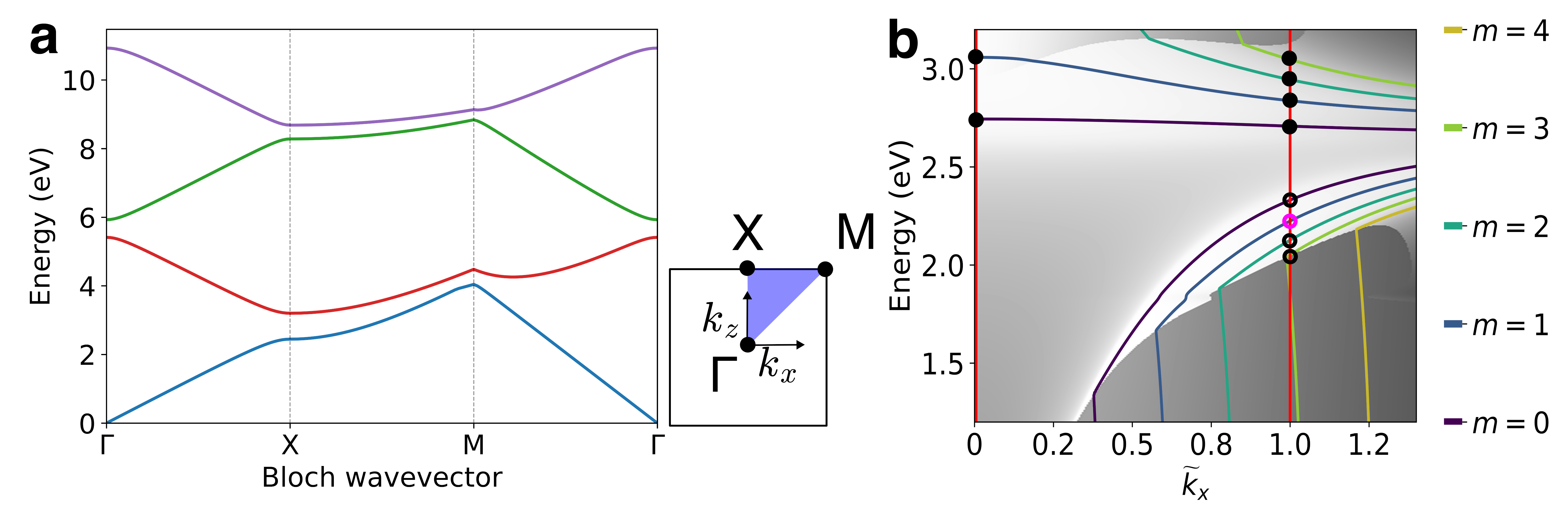}
	\caption{\small Infinite periodic multilayers. {\bf(a)}  Photonic bandstructure for  \ce{TiO2}/\ce{SiO2} photonic crystal with equal layer thicknesses of 60~nm,  calculated within the non-dispersive permittivity approximation (i.e., $\epsilon_1=constant$, $\epsilon_2=0$); inset represents the 2D Bruillouin zone. {\bf(b)} Complex-$\mathbf{k}$ bandstructure for a Ag/\ce{TiO2}  metamaterial with equal filling factor and period thickness of 50~nm, obtained from complex frequency-dependent dielectric function.  Red vertical lines indicate the fundamental ($\tilde{k}_x=0$) and the first ($\tilde{k}_x=1$) harmonic contributions associated with top grating of periodicity $\Lambda_x=100$~nm. Colored curves correspond to the first five  $k_z$ resonance values ($m = {0-4}$) for N = 6 metal/dielectric bilayers. Filled and open circles denote type-I and type-II VPP modes, respectively. The pink circle corresponds to the specific VPP resonance highlighted in Figure~\ref{fig:fig4}c and detailed in Figure~\ref{fig:fig6}b.}
	\label{fig:fig7}
\end{figure*}

\subsection{Electromagnetic fields}\label{sec:fields}
The \texttt{fields} functionality reconstructs the complete real-space EM fields. It combines the scattering-matrix solution with the eigenmodes of each layer, yielding one-dimensional field profiles along the stacking direction and two-dimensional field maps in the $x-z$ plane. This capability complements the optical spectra discussed in the previous sections and provides direct visualization of the EM states sustained by different classes of layered materials. To this end, we evaluated the EM field profiles for the three reference systems analyzed in the previous section: a \ce{TiO2}/\ce{SiO2} photonic crystal, an Ag/\ce{TiO2} HMM, and an air/Ag interface. For consistency, we retained the same structural parameters specified above, including grating. TM polarization was considered for all investigated systems. Specifically, we calculated the spatial distribution of the magnetic field associated with the selected EM resonances marked by open circles in the RTA spectra of Figures~\ref{fig:fig4} and~\ref{fig:fig5}, assumed as prototypical examples. The results are summarized in Figure~\ref{fig:fig6}.

For the photonic crystal (Figure~\ref{fig:fig6}a), the selected EM mode corresponds to an extended, traveling wave that is uniformly distributed along the $x$-direction. This wave successfully penetrates and propagates through the entire multilayer stack, in agreement with the prominent transmittance peak observed in Figure~\ref{fig:fig4}b.

For the Ag/\ce{TiO2}  metamaterial (Figure~\ref{fig:fig6}b), the selected resonance corresponds to a VPP mode that propagates throughout the multilayer as a collective polaronic excitation. The zero-intensity value of the magnetic field profile at the exit-surface/air interface confirms that this mode is unable to radiate out of the multilayer stack, which is in agreement with the vanishing transmittance observed in Figure~\ref{fig:fig4}b. Furthermore, the number of nodes in the 2D spatial maps of the magnetic field (bottom panels) allows for the classification of VPPs into distinct orders, providing a direct link to the experimental reflectance analysis. The maximum VPP order is determined by the number and thickness of the layers composing the metamaterial, Eq.~(\ref{bloch}), and sets the upper limit on the total number of VPP modes that can be excited within a fixed stack geometry.
Here, the EM state exhibits a single node, identifying it as the first-order ($1^{st}$) VPP mode. The bandstructure tools can be used to design and predict the optimal geometrical parameters governing the EM excitations supported by metamaterials (see Section~\ref{sec:cbands}).

Finally, for the air/Ag interface (Figure~\ref{fig:fig6}c), the magnetic field profile exhibits the characteristic behavior of a surface-bound EM mode. The field decays exponentially along the $z$-direction - vanishing more rapidly within the metallic layer than in the air region - while oscillating in-plane along the $x$-direction, precisely as expected for a SPP excitation.

\subsection{Photonic band analysis}
\label{sec:cbands}
To demonstrate the capabilities of the \texttt{bands} functionality of the \emerald~code, we considered two systems belonging to the same class of materials discussed above: a \ce{TiO2}/\ce{SiO2} photonic crystal and a Ag/\ce{TiO2}  hyperbolic metamaterial. In contrast to the previous finite-stack configurations, we now model infinite periodic structures. We thus calculate the PBS of the periodically repeated multilayers, which provides the overall distribution of the EM modes of the system (i.e., Bloch states).

First, we calculated the photonic bandstructure for the photonic crystal within the non-dispersive permittivity approximation    (i.e., $\epsilon_1=constant$, $\epsilon_2=0$).
Figure~\ref{fig:fig7}a shows the lowest-energy transverse magnetic (TM) photonic modes as a function of the wavevector {\bf k}
along a closed path in the $k_y=0$ plane of the Brillouin zone.
We identify a complete photonic bandgap of $\sim 0.6$~eV at
$\Gamma$, along with partial gaps at the high-symmetry points. The emergence of the bandgap results from the dielectric contrast between the two constituent materials, which leads to Bragg scattering at the Brillouin zone boundaries. The red and green bands represent the so-called \emph{dielectric} and \emph{air} modes which concentrate the EM energy in the regions with higher \ce{TiO2} and lower  \ce{SiO2} permittivity, respectively.

As a second example, we calculated the complex-$\mathbf{k}$ photonic bandstructure Ag/\ce{TiO2}  hyperbolic metamaterial obtained from complex frequency-dependent dielectric function. In Figure~\ref{fig:fig7}b, the pixel colors in the grayscale spectrum represent the magnitude of the imaginary part of the complex Bloch wavevector $k_z$ along the stacking direction  as a function of the photon energy $E$ and of the in-plane $k_x$ component of the wavevector. In this map, lighter regions correspond to low attenuation, where EM waves propagate with minimal loss; darker regions indicate strong attenuation associated with large imaginary wavevector components.

While the PBS plot provides continuous dispersion relations for the EM modes within an infinite multilayer, practical applications typically require identifying which specific modes can be excited in realistic systems. In such finite configurations, both the exact number of layers and the coupling mechanism are uniquely determined during the fabrication process.
To this end, we exploit the boundary matching tools implemented in the \texttt{bands} functionality described above. The in-plane periodicity induced by the grating selects specific (scaled) $\tilde{k}_x$ wavevectors, which are defined by diffraction harmonics and represented as vertical red lines in Figure~\ref{fig:fig7}b. Concurrently, the finite number of metal/dielectric bilayers restricts the allowed perpendicular $k_z$ components to discrete values, as expressed by Eq.~(\ref{bloch}). The colored curved lines in Figure~\ref{fig:fig7}b depict the isolines corresponding to the first five resonance $k_z$ values for a multilayer consisting of N=6 bilayers. The mode index $m$ (ranging from $0$ to $N-1$) indicates the number of nodes in the EM field within the metamaterial, thereby establishing a direct correlation between the resonance $k_z$ values and the VPP orders discussed previously. The intersections between the vertical red lines and the colored isolines correspond to the only physical EM states that can be excited within the metamaterial. Here, filled and open circles denote type-I and type-II VPP modes, respectively, while the pink circle identifies the specific VPP resonance highlighted in Figure~\ref{fig:fig4}c and detailed in Figure~\ref{fig:fig6}b.

\section{Conclusion}
\emerald~(\underline{E}lectro\underline{M}agnetic modes \underline{E}nginee\underline{R}ing in \underline{A}dvanced \underline{L}ayere\underline{D} materials) provides a unified, open-source framework for quantitative electromagnetic (EM) simulations of layered and periodically structured (meta)materials, integrating an efficient scattering-matrix implementation for finite stacks with rigorous coupled-wave analysis (RCWA), plane-wave expansion (PWE) tools for Bloch-modes and photonic band-structure (PBS) analysis. By combining these complementary formalisms within a single, modular and object-oriented workflow, the code bridges realistic device optics---where spectra and near fields are governed by interference, finite thickness, and boundary conditions---with the dispersion properties of the corresponding ideal periodic media.

\emerald~directly connects EM modelling with the underlying materials response.
Optical simulations are grounded on the electronic and optical properties of the constituent materials, which can be provided from experimental
measurements, literature data, or obtained from first-principles calculations, enabling predictive studies across dispersive, lossy,
and anisotropic platforms. Within the RCWA/PWE formulation, the Fourier-space treatment of
Maxwell's  equations provides a systematic route to handle laterally patterned couplers and periodic media,
while the scattering-matrix approach ensures robust simulations of multilayers with many periods,
strong impedance mismatch, or evanescent components.
Together, these complementary approaches allow for a direct comparison with experimental spectra and a clear physical interpretation of the supported EM modes.

Through three prototypical representatives of relevant classes of multilayers, we have shown how the joint use of {\bf i.} finite-stack optical spectra and field profiles and {\bf ii.} PBS and Bloch modes analyses enables an interpretable and design-oriented characterization of EM resonances and plasmon-polariton modes, in photonic crystals, and metamaterials. Overall, \emerald~supports the material--geometry co-design of advanced photonic and metamaterial devices, offering a practical tool to rationalize experiments, optimize geometry and coupling conditions, and engineer EM modes across dielectric photonic crystals, plasmonic multilayers, and hyperbolic metamaterials.

\section*{CRediT authorship contribution statement}
{\bf Stefano~Campanaro:} Writing – review \& editing, Writing – original
draft, Methodology, Code implementation, Tests and data generation, Conceptualization.
{\bf Luca~Bursi:} Writing – review \& editing, Writing – original
draft, Validation, Methodology, Conceptualization.
{\bf Nicholas~Anderson:} Writing – review \& editing,
Writing – original draft, Validation.
{\bf Stefano~Curtarolo:} Writing – review \& editing,
Writing – original draft, Supervision, Funding acquisition.
{\bf Arrigo~Calzolari:} Writing – review \& editing,
Writing – original draft, Validation, Supervision, Funding acquisition, Conceptualization.

\section*{Declaration of competing interest}
The authors declare that they have no known competing financial interests or personal relationships that could have appeared to influence the work reported in this paper.

\section*{Acknowledgements}
L.B. acknowledges financial support from PNRR MUR project ECS\_00000033\_ECOSISTER funded by the European Union-NextGenerationEU.
A.C. acknowledges the National Centre for HPC, Big Data and Quantum Computing (ICSC), funded under the National Recovery and Resilience Plan (NRRP), Mission 04 Component 2 Investment 1.4, NextGenerationEU, Award Number:CN00000013.
S.Cur. acknowledges support by the Office of Naval Research under grants N00014-23-1-2615 and N00014-24-1-2768, and by the DoD High Performance Computing Modernization Program (Frontier).
S.Cur. also thanks Auro Scientific, LLC for computational support.
The authors thank Drs. Simon Divilov, Hagen Eckert, Paolo De Angelis, Michael Mehl, Xiomara Campilongo and Douglas E. Wolfe for useful discussions.

\section*{Data and Code Availability}
Data and  source code of \emerald~are publicly available at \href{http://github.com/aflow-org/emerald}{github.com/aflow-org/emerald}.


\begin{thebibliography}{10}
\expandafter\ifx\csname urlstyle\endcsname\relax
  \providecommand{\doi}[1]{doi:\discretionary{}{}{}#1}\else
  \providecommand{\doi}{doi:\discretionary{}{}{}\begingroup
  \urlstyle{rm}\Url}\fi
\providecommand{\selectlanguage}[1]{\relax}
\providecommand{\bibAnnoteFile}[1]{%
  \IfFileExists{#1}{\begin{quotation}\noindent\textsc{Key:} #1\\
  \textsc{Annotation:}\ \input{#1}\end{quotation}}{}}
\providecommand{\bibAnnote}[2]{%
  \begin{quotation}\noindent\textsc{Key:} #1\\
  \textsc{Annotation:}\ #2\end{quotation}}

\bibitem{Shamim2024}
S.~Shamim, A.~S. Mohsin, M.~M. Rahman, and M.~B.~H. Bhuian, \emph{Recent
  advances in the metamaterial and metasurface-based biosensor in the
  gigahertz, terahertz, and optical frequency domains}, Heliyon \textbf{10},
  e33272 (2024).
\input{Shamim2024}

\bibitem{Lee2024}
D.~Lee, W.~W. Chen, L.~Wang, Y.-C. Chan, and W.~Chen, \emph{Data-driven design
  for metamaterials and multiscale systems: A review}, Adv. Mater. \textbf{36},
  2305254 (2024).
\input{Lee2024}

\bibitem{moradi2023}
A.~Moradi, \emph{{Theory of Electrostatic Waves in Hyperbolic Metamaterials}},
  \emph{Springer Series in Optical Sciences}, vol. 245 (Springer, Cham, 2023),
  \doi{10.1007/978-3-031-48596-1}.
\input{moradi2023}

\bibitem{vpp_hmm}
S.~V. Zhukovsky, O.~Kidwai, and J.~E. Sipe, \emph{Physical nature of volume
  plasmon polaritons in hyperbolic metamaterials}, Opt. Exp. \textbf{21},
  14982--14987 (2013).
\input{vpp_hmm}

\bibitem{Joannopoulos2008}
J.~D. Joannopoulos, S.~G. Johnson, J.~N. Winn, and R.~D. Meade, \emph{Photonic
  Crystals: Molding the Flow of Light} (Princeton University Press, Princeton,
  NJ, 2008), 2nd edn., \doi{10.1515/9781400828241}.
\input{Joannopoulos2008}

\bibitem{APELL199797}
P.~Apell and O.~Hunderi, \emph{Optical properties of superlattices}, in
  \emph{Handbook of Optical Constants of Solids}, edited by E.~D. Palik
  (Academic Press, Burlington, 1997), pp. 97--124,
  \doi{10.1016/B978-0-08-055630-7.50008-0}.
\input{APELL199797}

\bibitem{Smolyaninov2018}
I.~I. Smolyaninov, \emph{Hyperbolic Metamaterials}, 2053--2571 (Morgan \&
  Claypool Publishers, IOP Publishing Limited, 2018),
  \doi{10.1088/978-1-6817-4565-7}.
\input{Smolyaninov2018}

\bibitem{pod_hmm}
A.~Poddubny, I.~Iorsh, P.~Belov, and Y.~Kivshar, \emph{Hyperbolic
  metamaterials}, Nat. Photonics \textbf{7}, 948--957 (2013).
\input{pod_hmm}

\bibitem{excitation2014}
K.~V. Sreekanth, A.~De~Luca, and G.~Strangi, \emph{Excitation of volume plasmon
  polaritons in metal-dielectric metamaterials using 1{D} and 2{D} diffraction
  gratings}, J. Opt. \textbf{16}, 105103 (2014).
\input{excitation2014}

\bibitem{em_multilayer}
O.~Kidwai, S.~V., and J.~E. Sipe, \emph{Effective-medium approach to planar
  multilayer hyperbolic metamaterials: Strengths and limitations}, Phys. Rev. A
  \textbf{85}, 053842 (2012).
\input{em_multilayer}

\bibitem{smm1}
N.~P.~K. Cotter, T.~W. Preist, and J.~R. Sambles, \emph{Scattering-matrix
  approach to multilayer diffraction}, J. Opt. Soc. Am. A \textbf{12},
  1097--1103 (1995).
\input{smm1}

\bibitem{smm2}
R.~C. Rumpf, \emph{Improved formulation of scattering matrices for
  semi-analytical methods that is consistent with convention}, Prog.
  Electromagn. Res. B \textbf{35}, 241--261 (2011).
\input{smm2}

\bibitem{DeBruijn2025}
Y.~De~Bruijn and E.~O. Hiltunen, \emph{Complex band structure for subwavelength
  evanescent waves}, Stud. Appl. Math. \textbf{154}, e70022 (2025).
\input{DeBruijn2025}

\bibitem{Li2003}
Z.-Y. Li and L.-L. Lin, \emph{Photonic band structures solved by a
  plane-wave-based transfer-matrix method}, Phys. Rev. E \textbf{67}, 046607
  (2003).
\input{Li2003}

\bibitem{born-wolf1999}
M.~Born, E.~Wolf, A.~B. Bhatia, P.~C. Clemmow, D.~Gabor, A.~R. Stokes, A.~M.
  Taylor, P.~A. Wayman, and W.~L. Wilcock, \emph{Principles of Optics:
  Electromagnetic Theory of Propagation, Interference and Diffraction of Light}
  (Cambridge University Press, 1999), 7 edn., \doi{10.1017/CBO9781139644181}.
\input{born-wolf1999}

\bibitem{berreman}
D.~W. Berreman, \emph{Optics in stratified and anisotropic media: 4$\times$
  4-matrix formulation}, J. Opt. Soc. Am. A \textbf{62}, 502--510 (1972).
\input{berreman}

\bibitem{Moharam1981}
M.~G. Moharam and T.~K. Gaylord, \emph{Rigorous coupled-wave analysis of
  planar-grating diffraction}, J. Opt. Soc. Am. A \textbf{71}, 811--818 (1981).
\input{Moharam1981}

\bibitem{HMMs_arXiv_2026}
S.~Campanaro, L.~Bursi, S.~Curtarolo, and A.~Calzolari, \emph{Terahertz volume
  plasmon-polariton modulation in all-dielectric hyperbolic metamaterials},
  Adv. Opt. Mater. \textbf{14}, e02680 (2026).
\input{HMMs_arXiv_2026}

\bibitem{nmatHT}
S.~Curtarolo, G.~L.~W. Hart, M.~{Buongiorno Nardelli}, N.~Mingo, S.~Sanvito,
  and O.~Levy, \emph{The high-throughput highway to computational materials
  design}, Nat. Mater. \textbf{12}, 191--201 (2013).
\input{nmatHT}

\bibitem{aflow4}
S.~Divilov, H.~Eckert, S.~D. Thiel, S.~D. Griesemer, R.~Friedrich, N.~H.
  Anderson, M.~J. Mehl, D.~Hicks, M.~Esters, N.~Hotz, X.~Campilongo,
  A.~Calzolari, and S.~Curtarolo, \emph{AFLOW4: Heading Toward Disorder}, High
  Entropy Alloys Mater. \textbf{3}, 178--187 (2025).
\input{aflow4}

\bibitem{curtarolo:art115}
C.~Toher, C.~Oses, J.~J. Plata, D.~Hicks, F.~Rose, O.~Levy, M.~{de Jong},
  M.~Asta, M.~Fornari, M.~{Buongiorno Nardelli}, and S.~Curtarolo,
  \emph{Combining the {AFLOW} {GIBBS} and elastic libraries to efficiently and
  robustly screen thermomechanical properties of solids}, Phys. Rev. Mater.
  \textbf{1}, 015401 (2017).
\input{curtarolo:art115}

\bibitem{curtarolo:art187}
{A. Calzolari}, {C. Oses}, {C. Toher}, {M. Esters}, {X. Campilongo}, {S.~P.
  Stepanoff}, {D.~E. Wolfe}, and {S. Curtarolo}, \emph{{Plasmonic high-entropy
  carbides}}, Nat. Commun. \textbf{13}, 5993 (2022).
\input{curtarolo:art187}

\bibitem{curtarolo:art223}
S.~Divilov, S.~D. Griesemer, R.~C. Koennecker, M.~J. Ammendola, A.~C. Zettel,
  H.~Eckert, J.~R. Shallenberger, X.~Campilongo, W.~G. Fahrenholtz,
  A.~Calzolari, D.~E. Wolfe, and S.~Curtarolo, \emph{Variable-temperature
  plasmonic high-entropy carbides}, High Entropy Alloys Mater. \textbf{3},
  273--284 (2025).
\input{curtarolo:art223}

\bibitem{deed}
S.~Divilov, H.~Eckert, D.~Hicks, C.~Oses, C.~Toher, R.~Friedrich, M.~Esters,
  M.~J. Mehl, A.~C. Zettel, Y.~Lederer, E.~Zurek, J.-P. Maria, D.~W. Brenner,
  X.~Campilongo, S.~Filipovic, W.~G. Fahrenholtz, C.~J. Ryan, C.~M. DeSalle,
  R.~J. Crealese, D.~E. Wolfe, A.~Calzolari, and S.~Curtarolo,
  \emph{{Disordered enthalpy-entropy descriptor for high-entropy ceramics
  discovery}}, Nature \textbf{625}, 66--73 (2024).
\input{deed}

\bibitem{Moharam95}
M.~G. Moharam, E.~B. Grann, D.~A. Pommet, and T.~K. Gaylord, \emph{Formulation
  for stable and efficient implementation of the rigorous coupled-wave analysis
  of binary gratings}, J. Opt. Soc. Am. A \textbf{12}, 1068--1076 (1995).
\input{Moharam95}

\bibitem{Li96}
L.~Li, \emph{Use of Fourier series in the analysis of discontinuous periodic
  structures}, J. Opt. Soc. Am. A \textbf{13}, 1870--1876 (1996).
\input{Li96}

\bibitem{FigotinVitebskiy2006}
A.~Figotin and I.~Vitebskiy, \emph{Electromagnetic unidirectionality in
  magnetic photonic crystals}, Phys. Rev. B \textbf{74}, 155128 (2006).
\input{FigotinVitebskiy2006}

\bibitem{Rybin2017}
M.~V. Rybin and M.~F. Limonov, \emph{Inverse dispersion method for calculation
  of complex photonic band diagrams}, Phys. Rev. A \textbf{96}, 013839 (2017).
\input{Rybin2017}

\bibitem{JohnsonJoannopoulos2001}
S.~G. Johnson and J.~D. Joannopoulos, \emph{{B}lock-iterative frequency-domain
  methods for {M}axwell's equations in a planewave basis}, Opt. Express
  \textbf{8}, 173--190 (2001).
\input{JohnsonJoannopoulos2001}

\bibitem{redheffer1959}
R.~Redheffer, \emph{Inequalities for a matrix {R}iccati equation}, Indiana
  Univ. Math. J. pp. 349--367 (1959).
\input{redheffer1959}

\bibitem{qe}
P.~Giannozzi, S.~Baroni, N.~Bonini, M.~Calandra, R.~Car, C.~Cavazzoni,
  D.~Ceresoli, G.~L. Chiarotti, M.~Cococcioni, I.~Dabo, A.~D. Corso,
  S.~de~Gironcoli, S.~Fabris, G.~Fratesi, R.~Gebauer, U.~Gerstmann,
  C.~Gougoussis, A.~Kokalj, M.~Lazzeri, L.~Martin-Samos, N.~Marzari, F.~Mauri,
  R.~Mazzarello, S.~Paolini, A.~Pasquarello, L.~Paulatto, C.~Sbraccia,
  S.~Scandolo, G.~Sclauzero, A.~P. Seitsonen, A.~Smogunov, P.~Umari, and R.~M.
  Wentzcovitch, \emph{QUANTUM ESPRESSO: a modular and open-source software
  project for quantum simulations of materials}, J. Phys.: Condens. Matter
  \textbf{21}, 395502 (2009).
\input{qe}

\bibitem{TiO2}
D.~Franta, D.~Necas, I.~Ohl{\'i}dal, and A.~Giglia, \emph{{Dispersion model for
  optical thin films applicable in wide spectral range}}, in \emph{Optical
  Systems Design 2015: Optical Fabrication, Testing, and Metrology V}, edited
  by A.~Duparr{\'e} and R.~Geyl, International Society for Optics and Photonics
  (SPIE, 2015), vol. 9628, p. 96281U, \doi{10.1117/12.2190104}.
\input{TiO2}

\bibitem{SiO2}
D.~Franta, D.~Necas, I.~Ohl{\'i}dal, and A.~Giglia, \emph{{Optical
  characterization of SiO2 thin films using universal dispersion model over
  wide spectral range}}, in \emph{Optical Micro- and Nanometrology VI}, edited
  by C.~Gorecki, A.~K. Asundi, and W.~Osten, International Society for Optics
  and Photonics (SPIE, 2016), vol. 9890, p. 989014, \doi{10.1117/12.2227580}.
\input{SiO2}

\bibitem{ferrari}
L.~Ferrari, C.~Wu, D.~Lepage, X.~Zhang, and Z.~Liu, \emph{Hyperbolic
  metamaterials and their applications}, Prog. Quantum Electron. \textbf{40},
  1--40 (2015).
\input{ferrari}

\bibitem{Ag-CIESIELSKI2017}
A.~Ciesielski, L.~Skowronski, M.~Trzcinski, and T.~Szoplik, \emph{Controlling
  the optical parameters of self-assembled silver films with wetting layers and
  annealing}, Appl. Surf. Sci. \textbf{421}, 349--356 (2017).
\input{Ag-CIESIELSKI2017}

\bibitem{takagi2017}
K.~Takagi, S.~V. Nair, R.~Watanabe, K.~Seto, T.~Kobayashi, and E.~Tokunaga,
  \emph{Surface plasmon polariton resonance of gold, silver, and copper studied
  in the kretschmann geometry: Dependence on wavelength, angle of incidence,
  and film thickness}, J. Phys. Soc. Japan \textbf{86}, 124721 (2017).
\input{takagi2017}

\bibitem{Ag-Johnson}
P.~B. Johnson and R.~W. Christy, \emph{Optical Constants of the Noble Metals},
  Phys. Rev. B \textbf{6}, 4370--4379 (1972).
\input{Ag-Johnson}

\end{thebibliography}
\end{document}